Review

# Antimatter Gravity Experiments, the Astronomical Challenges to ΛCDM Cosmology and the Quantum Vacuum as a Possible Source of Gravity in the Universe

Dragan Slavkov Hajdukovic

INFI, Cetinje, Montenegro
dragan.hajdukovic@alumni.cern
dragan.hajdukovic@gmail.com

**Abstract**: This review is motivated by the first result of the ALPHA-g experiment at CERN, which indicates that atoms and anti-atoms have different gravitational charges; according to measurements, the gravitational acceleration of anti-atoms is only 0.75 of that of ordinary matter. If confirmed by more precise measurements, this will greatly increase the plausibility of the emerging cosmological model, which is based on the working hypothesis that quantum vacuum fluctuations are virtual gravitational dipoles; a hypothesis that opens up the possibility that the quantum vacuum is a major source of gravity in the universe (which could eventually eliminate the need for the hypothetical dark matter and dark energy). This laboratory challenge to general relativity and ΛCDM cosmology is complemented by astronomical challenges (the Hubble tension, very fast initial growth of structures in the Universe, dark energy deviating from the cosmological constant...). The intriguing question is: do the antimatter gravity experiments at CERN and the recent astronomical observations point to the same (highly unexpected) new physics? With this question in mind, we briefly review the antimatter gravity experiments at CERN and elsewhere, together with the major astronomical challenges and the emerging Quantum Vacuum cosmology, which seems to be compatible with the ALPHA-g result and with the preliminary astronomical challenges to ΛCDM cosmology. The laboratory and astronomical challenges have suddenly taken us into terra incognita, where we need absolutely unprecedented imagination and open-minded thinking.

Keywords：antimatter gravity experiments; gravitational dipoles; quantum vacuum cosmology; evolving dark energy; Hubble tension; accelerated formation of galaxies

## 1. Introduction

Contemporary physics has two cornerstones, General relativity and the Standard Model of particles and fields.

General relativity is the wrong name for our best theory of gravity, which reduces to Newtonian gravitation in the limit of weak fields and slow motions.

The Standard Model of particles and fields is our best theory of fundamental particles and their interactions. According to the Standard Model of particles and fields, *everything* in the Universe (i.e. the matter-energy content of the Universe) is made up of apparently structureless, point-like fermions (quarks and leptons, antiquarks and antileptons) interacting through the exchange of gauge bosons (gluons for the strong interaction; photons for the electromagnetic interaction; $Z^0, W^+$ and $W^-$ bosons for the weak interaction). This well-established content of the Universe can be called "Standard Model matter" (hereafter referred to as SM matter) and must be clearly distinguished from the non-established, i.e. hypothetical, content of the Universe, such as dark matter and dark energy.

The Standard Model of particles and fields is strictly limited to strong, electromagnetic and weak interactions; gravitational interactions are completely excluded from the Standard Model. We don't know, and can only guess, what the gravitational properties of quarks and antiquarks, leptons and antileptons, and gauge bosons are. There are many questions we can't answer; for example, we don't know whether quarks and antiquarks (and similarly leptons and antileptons) are gravitationally indistinguishable. These and similar question are not crucial to Newtonian gravity (and, while in a more complex way to General relativity), in which the gravitational interactions between macroscopic bodies are determined by the masses of the bodies and, fortunately for the development of the theory, the interactions are independent of the internal (then completely unknown) atomic structure of the bodies. Although not important for Newtonian gravity and General relativity, knowing the gravitational properties of the fundamental fermions and bosons of the Standard Model is crucial for understanding the gravitational effects of the quantum vacuum in the Universe

The quantum vacuum is inherent part of the Standard Model of particles and fields. The key point is that the physical vacuum is not just empty space (nothingness). In quantum field theory, nothing is plenty [1].

If you are not familiar with the quantum vacuum, just think of it as a new state of matter-energy, radically different from the more familiar states of matter (solid, liquid, gas, ordinary plasma, quark-gluon plasma, ...), but as real as they are. You can imagine the quantum vacuum as an omnipresent fluid composed of quantum vacuum fluctuations, or, in a more popular wording, composed of virtual particle–antiparticle pairs with extremely short lifetimes (for example, the lifetime of a virtual electron–positron pair is of the order of $10^{-21}$ seconds). The short-lived "inhabitants" of the quantum vacuum are therefore virtual quark-antiquark pairs, virtual lepton-antilepton pairs, but also virtual gauge bosons, which perpetually appear and disappear (as is allowed by the time-energy uncertainty relation). Note that the quantum vacuum differs from all other states of matter in that it contains equal amounts of matter and antimatter; in a quantum vacuum fluctuation, a particle always appears paired with its antiparticle, which is completely different from the mystery of the matter-antimatter asymmetry in the Universe.

In the study of the electromagnetic, strong and weak interactions, the quantum vacuum cannot be neglected; there are well-established non-gravitational interactions between the quantum vacuum and the matter immersed in it. There is a non-gravitational symbiosis between the quantum vacuum and the Standard Model matter immersed in it. Let us mention just three fascinating, illuminating and experimentally confirmed phenomena of this symbiosis.

First, the quantum vacuum has a permanent tiny impact (but impact!) on 'orbits' (i.e. energy levels) of electrons in atoms [1]; this phenomenon is known as the „Lamb shift".

Second, under the influence of a sufficiently strong, external electric field, quantum vacuum fluctuations can become polarized. A simple 'mental picture' of this phenomenon is as follows. A virtual electron–positron pair (i.e. two electric charges of the opposite sign) is actually a virtual electric dipole and, from the point of view of electrodynamics, a quantum vacuum is an 'ocean' of randomly oriented electric dipoles; a strong electric field can force (can impose) the alignment of these dipoles. Thus the electric polarization of the quantum vacuum is as real as the analogous polarization of a dielectric. Charged particles (electrons, positrons, protons, . . . ) create a *microscopic halo* of the polarized quantum vacuum around themselves; the effect of this halo is the 'screening' of the electric charge of the particle. Consequently, if you measure the electric charge of an electron outside its halo of polarized quantum vacuum, you will get the familiar constant value ($1.602 \times 10^{-19}$C); however, if you measure it inside the halo (where the screening is smaller), you will measure a larger electric charge [2].

Third, quantum vacuum fluctuations can be converted into real particles; we can create something from apparently nothing. In fact, more than a decade ago [3], the dynamical Casimir effect (i.e. the creation of photons from the quantum vacuum) was confirmed; poetically speaking, for the first time, we created light from darkness.

The immediate open question is whether there are also gravitational interactions between the quantum vacuum and the Standard Model matter immersed in it. Have we so far neglected the gravitational symbiosis between the quantum vacuum and Standard Model matter? This question cannot be answered without knowing the gravitational properties of the fundamental particles (quarks, antiquarks, leptons, antileptons and gauge bosons).

Whatever the answer is, the lesson we have learned from the cosmological constant problem [4] is that we are missing something very fundamental in our understanding of gravity. The essence of the cosmological constant problem is that, if our current understanding of gravity is extrapolated to fundamental particles, the quantum vacuum must produce gravitational effects many orders of magnitude larger than the empirical evidence allows. More explicitly, the virtual mass-energy density of the quantum vacuum (theoretically predicted by the Standard Model of particles and fields) is many orders of magnitude greater than a possible gravitational charge density of the quantum vacuum allowed by astronomical observations. Apparently the quantum vacuum does not respect our prescribed truth that mass–energy and gravitational charge must be the same quantity.

The main question is whether contemporary physics (General relativity and the Standard Model of particles and fields) is sufficient to explain astronomical observations on scales much larger than our Solar System. In short, astronomical observations of galaxies, galaxy clusters and the entire visible Universe are challenging contemporary physics with a number of phenomena that are a complete surprise and a complete mystery (to the extent that we may be in the greatest crisis in the history of physics). The main examples of these challenging phenomena are:

- In galaxies and galaxy clusters, *the gravitational field is much stronger* than it should be according to the amount of Standard Model matter and our theory of gravity.
- T*he expansion of our Universe is accelerating*; contrary to the common sense expectation that gravity should slow the rate of expansion.
- *Our Universe is dominated by matter*; apparently something in the primordial Universe forced the matter-antimatter asymmetry.

There are two main paradigms as a scientific response to the above problems. Common to both paradigms is that, because of the cosmological constant problem, the quantum vacuum is not directly included as a source of gravity in the Universe.

The first and dominant paradigm is the ΛCDM Cosmology in which General relativity is completely trusted, but in addition to Standard Model matter, a number of ad hoc hypotheses are invoked; in, fact, one ad hoc hypothesis for each mystery.  To explain the observations, the ΛCDM cosmology has filled the Universe with hypothetical dark matter and dark energy while in a tiny fraction of second after the Big Bang, there is a mysterious inflationary field (causing a monstrous initial exponential expansion of the Universe at a speed many orders of magnitude greater than the speed of light) and a huge CP violation of unknown nature that caused domination of matter over antimatter in the Universe. And after all these hypotheses, we still have no plausible idea of what underlies the cosmological constant problem. We do not know if our hypotheses are correct, or if they are just good imitations of something we do not understand. In any case, the current theoretical thinking (forced by the observed mysteries) is a departure from the traditional elegance, simplicity and beauty of theoretical physics.

The second paradigm accepts Standard Model matter as the only content of the Universe, and attempts to explain the observed phenomena by an appropriate modification of gravity.

Beyond these two dominant paradigms, there are a number of alternative cosmologies (see, for example, a review published this year [5]). Alternative cosmologies are based on a wide range of very different ideas, from entropic cosmology [6] to f(Q) gravity [7]. As the author of the review [5] rightly points out: To a greater or lesser extent, all these alternative models suffer from a lack of development compared to the standard ΛCDM, because many of these alternative ideas are in the hands of very few individuals who cannot produce results at the same speed as the thousands of researchers working on the standard model.

As noted above, the present review is motivated by the preliminary results of antimatter gravity experiments; consequently, we focus only on alternative cosmologies that depend to some extent on the gravitational properties of antimatter. Three very different theories of this type have emerged in the early 21st century, all exploring the intriguing possibility that the gravitational properties of antimatter may be the cause of phenomena usually attributed to dark matter and dark energy.

Two of these theories, the Lattice Universe [8, 9] and the Dirac-Milne Universe [10-12] are based on the common assumption that the universe contains equal amounts of matter and antimatter (so there is no problem of matter-antimatter asymmetry), but they attribute two radically different (in fact, incompatible) gravitational properties to antiatoms. In the Lattice Universe, the gravitational attraction between antiatoms is the same as the gravitational attraction between atoms, but there is mutual gravitational repulsion between atoms and antiatoms. In the Dirac-Milne cosmology, of course, there is gravitational attraction between atoms, but there is a postulated gravitational repulsion between antiatoms (hence antiatoms are always a gas with no potential to form structures like antistars and antigalaxies); moreover, atoms gravitationally attract antiatoms, but antiatoms gravitationally repel atoms. In the Dirac-Milne cosmology, this complex set of assumed gravitational interactions inevitably leads to a bimetric theory of gravity. Appendix A provides some additional information on these two alternative cosmologies.

The third paradigm [13-18], which can be called Quantum Vacuum Cosmology (QV Cosmology), assumes a matter-dominated universe, making it radically different and incompatible with the Lattice Universe and Dirac-Milne Cosmology, which assume equal amounts of matter and antimatter in the universe. The immediate question is, if there is no antimatter in the universe, how can the gravitational properties of non-existent antimatter affect the evolution of the universe? In principle, the obvious (but not easy to see) answer to this question is the quantum vacuum. QV cosmology is based on the seemingly wild working hypothesis that at least some of the quantum-vacuum fluctuations behave as virtual gravitational dipoles; the quantum-vacuum "enriched" with virtual gravitational dipoles can be polarized by the intruding Standard Model matter (baryonic matter in astronomical jargon); the polarized quantum-vacuum acts as a source of gravity that can explain phenomena usually attributed to dark matter and dark energy and/or modifications of gravity.

It should be emphasized that QV cosmology is limited to studying the consequences of the possible existence of virtual gravitational dipoles in the quantum vacuum, without addressing the nature of these dipoles; this is a pragmatic approach, since at this stage of knowledge we are far from being able to reveal the physical nature of dipoles. For various reasons (such as the cosmological constant problem), it is preferable that the fundamental particles and antiparticles of the Standard Model have gravitational charges of opposite signs. However, the gravitational polarization of the quantum vacuum can also work with gravitational dipoles that are not associated with antimatter (see [19, 20] and Section 5) and this is the additional strength of QV cosmology.

The surprising preliminary result of the ALPHA-g experiment at CERN [21], indicating that atoms and antiatoms have different gravitational charges, *suggests the possibility that gravitational dipoles may exist at the level of the fundamental particles of the Standard Model of particles and fields*, and it is a good news for QV Cosmology. Moreover, it is obvious that the hypothesis of the existence of virtual gravitational dipoles in the quantum vacuum is not in any conflict with the Standard Model.

Surprisingly, the existence of virtual gravitational dipoles doesn't require any modification of General relativity either.

General relativity is the gravitational theory of matter immersed in a gravitationally featureless vacuum (i.e. classical vacuum), assuming the validity of the Weak Equivalence Principle. Thus, whatever the gravitational properties of antiatoms, General relativity remains valid in our Universe, which is dominated by atomic matter. Similarly, if we trust CPT symmetry, General relativity will be valid in a Universe dominated by antiatoms. If the ALPHA-g result is correct, then General relativity can only be challenged in a Universe that somehow contains both matter and antimatter. In addition, we will see in Section 3 that the quantum vacuum can be treated in a way analogous to material media in electrodynamics, where electrodynamics in material media doesn't invalidate electrodynamics in the vacuum.

Section 2 of this review is devoted to antimatter gravity experiments. First, we present the preliminary result of the ALPHA-g experiment, which indicates that anti-atoms fall in the Earth's gravitational field with a lower acceleration than atoms. In the second part of this section it is shown (on the basis of the proton and antiproton mass decomposition in Quantum chromodynamics) that the ALPHA-g result suggests the opposite sign of the gravitational charge of quarks and antiquarks. The third part of the section is a brief review of competing antimatter gravity experiments with anti-atoms at CERN, together with planned experiments in other laboratories in the lepton sector of the Standard Model (with positronium and muonium, which can clearly show whether leptons and antileptons have the opposite sign of the gravitational charge).

Section 3 is an overview of Quantum Vacuum Cosmology, an emerging model of the Universe that will attract much more attention if the preliminary ALPHA-g suggestion of the existence of gravitational dipoles is confirmed.

Section 4 is a brief review of some recent astronomical challenges to ΛCDM cosmology, suggesting that these challenges may in fact point to Quantum Vacuum Cosmology.

Section 5 gives an overview of the spectrum of different possibilities for the existence of gravitational dipoles, which (surprisingly at first sight) can also exist within the framework of General relativity and its most natural extensions (as General relativity with spin and torsion). In fact, this section shows that the outcome of the antimatter gravity experiments, while very important for QV cosmology, is not decisive.

An outlook is given in section 6.

Appendix A presents alternative cosmologies with a matter-antimatter symmetric universe.

Appendix B complements the review by pointing out analytical solutions, quantitative successes and open questions (which are at the same time opportunities for testing) of QV cosmology. The main purpose of the Appendix B is to help the reader to understand the problems and successes of QV cosmology without reading the original references.

## 1.1 QV Cosmology and the Emerging Empirical Evidence

It should be stressed that we must be very careful in interpreting the relationship between QV cosmology and the emerging empirical evidence from antimatter gravity experiments (Section 2) and astronomical observations (Section 4).

If confirmed, the preliminary result that atoms fall with greater gravitational acceleration than anti-atoms, will provide a fascinating window into new physics. While the final outcome of antimatter gravity experiments is still uncertain, far from our laboratories (up to 13.5 billion light-years away), the James Webb Space Telescope (JWST) and other sophisticated instruments such as DESY have probably already detected signatures of new physics.

It seems that challenges to ΛCDM cosmology and signatures of new physics are arriving almost simultaneously, both from laboratories and from astronomical observations. The extremely important fundamental question is: do the antimatter gravity experiments and the recent astronomical observations point to the same (highly unexpected) new physics? In other words, do phenomena observed at the scale of a small laboratory and at the largest scale of the whole visible Universe have a common root?

Is it possible that (after some further development of the theory) QV cosmology could be such a common root; a common explanation for both laboratory and astronomical challenges to ΛCDM cosmology? The first point to understand is that it is absolutely premature to answer this question. While QV cosmology has some well-established predictions (presented in Section 3 and the Appendix), there are topics for which only plausibility arguments and "speculative imagination" are possible, instead of the rigorous scientific approach. We believe that the speculation (speculative imagination) used in some parts of this Review is valuable and necessary in the long process of scientific discovery, and can stimulate new areas of research, but it must be clear that it is nothing more than that, and is in no way a proof of QV cosmology.

The second point to understand is that antimatter gravity experiments *cannot prove* (or disprove) the QV cosmology, but they can increase (or decrease) its plausibility. For example, establishing that quark-antiquark and lepton-antilepton pairs are gravitational dipoles will not prove QV cosmology, but such a quantum leap from the belief that gravitational dipoles are impossible to empirical evidence for their existence will increase the plausibility of the working hypothesis that quantum vacuum fluctuations could be gravitational dipoles.

## 2. Laboratory Challenges to ΛCDM Cosmology

### 2.1. The Preliminary Result of the ALPHA-g Experiment at CERN

Recently, for the first time in human history, ALPHA-g experiment at CERN [21], measured the gravitational acceleration of antiatoms (in particular acceleration of antihydrogen) in the gravitational field of the Earth. More precisely, the best fit to the ALPHA-g measurements gives

$$a_{\bar{H}} = [0.75 \pm 0.13(\text{statistical } + \text{ systematic}) \pm 0.16 \text{ (simulation)}]g$$

for the local acceleration $a_{\bar{H}}$ of antihydrogen towards the Earth. The 'g' denotes the local acceleration of gravity, which, for matter, is about 9.81 m/s². The significance of this breakthrough was immediately recognised [22, 23].

In other words: $a_{\bar{H}} = 7.4^{+2.8}_{-2.8}\, m/s^2$, instead of $g = 9.81\, m/s^2$. Although we must be cautious (because the accuracy of this preliminary result is only about 20%), the difference is surprisingly large and cannot be ignored. Before the ALPHA-g experiment, the vast majority (perhaps more than 99% of scientists) were completely convinced that atoms and anti-atoms have the same gravitational charge; after the ALPHA-g experiment, this mainstream opinion is seriously challenged.

According to this preliminary result of the ALPHA-g experiment, atoms fall with greater acceleration than anti-atoms in the Earth's gravitational field. However, in the gravitational field of an anti-Earth, antiatoms should fall with greater acceleration than atoms, as a consequence of CPT (charge-parity-time) symmetry, which tells us that the antimatter-antimatter interaction should be the same as the matter-matter interaction, but is also supported by the mass decomposition of nucleons and antinucleons in Quantum chromodynamics (discussed in the next subsection). Consequently, the weak equivalence principle holds in both an atomic and an antiatomic universe, but is violated in a universe composed of comparable amounts of atoms and antiatoms. It seems that the ALPHA-g experiment tells us that Einstein was lucky that we live in a Universe that is not a mixture of matter and antimatter regions, because in such a Universe the weak equivalence principle and his General theory of relativity would be wrong.

Let us remember that the Weak Equivalence Principle is the oldest and most trusted principle in modern physics. Its roots go back to the time of Galileo and his discovery of the "universality of free fall" on Earth. A deeper understanding was achieved by Newton, who explained the "universality of free fall" as a consequence of the equivalence of inertial mass and gravitational mass (better said, the gravitational charge). Unlike many other principles that lose their relevance over time, the WEP has actually gained in importance and is now the cornerstone of Einstein's General theory of relativity and modern cosmology. It is amusing that physicists use the word "weak" for a principle that has had such a long and successful life, and that has been associated with such great names (Galileo, Newton and Einstein). It should rather be called "the principle of giants".

Are we on the way to discovering that the Weak Equivalence Principle (known to hold for the atomic matter of our universe) is violated at the level of the fundamental fermionic particles and antiparticles of the Standard Model of particles and fields?

2.2. How to Interpret the ALPHA-g Result

While the gravitational charges of atoms and antiatoms have the same sign, the difference in the magnitude of the charges may be the first indication that the gravitational charges of quarks and antiquarks have opposite signs.

The immediate fundamental question (recently initiated by Menary [24], who is a member of the ALPHA-g collaboration) is how to reconcile the Quantum Chromodynamics and a possible WEP violation. The second question is whether and how the challenge coming from CERN and the challenges coming from astronomical observations can have the same origin.

According to the Quantum chromodynamics a proton (and similarly a neutron) is a complex dynamical system composed of relativistic quarks and gluons. The quarks in a proton belong to two very different subsets. First, there are 3 valence quarks $uud$. Second, there are "sea quarks" (or quark condensate), which appear in short-lived virtual quark-antiquark pairs. In principle, the percentages of these three contributions to the proton mass can be determined from a non-perturbative analytical solution based on the QCD Lagrangean. In the absence of an analytical solution a powerful numerical method called „lattice QCD" is used.

There are different (two, three and four term) decompositions of the proton mass from the QCD energy-momentum tensor. In principle, a more detailed decomposition is preferable when trying to guess the gravitational properties of the antiproton, so in Table 1 we give relatively accurate [25, 26] four-term decomposition. According to [25, 26], the quark and gluon energy contributions are 32% and 36% respectively; the additional 9% comes from sea-quarks and 23% results from the anomalous gluon contribution (caused by the QCD trace anomaly). In Table 1, $M_{ge}$ and $M_{vqe}$ denote the mass contributions corresponding to the gluon and valence quark energy contributions, respectively; similarly, $M_{ag}$ and $M_{sq}$ are the mass contributions corresponding to the anomalous gluon and sea-quark contributions.

**Table 1.** Approximate contributions of the four sources to the proton and antiproton masses $M_p$ and $M_{\bar{p}}$; ($M_p = M_{\bar{p}}$), adapted from [15, 16].

| gluons | QCD trace anomaly (anomalous gluon contribution) | - valence quarks $uud$ for protons<br>- valence antiquarks $\bar{u}\bar{u}\bar{d}$ for antiprotons | sea-quarks |
|---|---|---|---|
| $M_{ge} = 0.36 M_p$ | $M_{ag} = 0.23 M_p$ | $M_{vqe} = 0.32 M_p$ | $M_{sq} = 0.09 M_p$ |

Note that Table 1 includes the "QCD trace anomaly", a purely quantum effect which, according to Hatta [27], comes exclusively from the gluon momentum energy tensor (while the quark mass term comes only from the quark energy tensor) and should be considered as an additional part of the gluon contribution (anomalous gluon contribution).

We know that the proton and the antiproton have the same mass ($M_p = M_{\bar{p}}$), and that the mass decomposition given in Table 1 also applies to the antiproton. The crucial difference is that for the proton the valence quarks are $uud$, whereas for the antiproton the corresponding valence antiquarks are $\bar{u}\bar{u}\bar{d}$.

If in the Earth's gravitational field (it would be the opposite in the anti-Earth gravitational field) antiprotons fall with less acceleration than protons, this implies that the gravitational mass (or rather the gravitational charge) of the antiproton is less than the gravitational charge of the proton. Based on the evidence presented in Table 1, Menary suggested that such a violation of the WEP could eventually be reconciled with QCD if the valence quarks in the proton and the valence antiquarks in the antiproton have the opposite sign of the gravitational charge (in other words, if quark-antiquark pairs are virtual gravitational dipoles). If this is the case, then the gravitational charge corresponding to the gluons must be positive in both protons and antiprotons.

However, things may be much more complex than Menary's initial suggestion. It is fascinating that perhaps in the next decade at the latest, antimatter gravity experiments (in both the quark and lepton sectors of the Standard Model) will reveal whether quark-antiquark and lepton-antilepton bound systems are gravitational dipoles, i.e. whether quarks and antiquarks and leptons and antileptons have opposite signs of gravitational charge. However, the potentially surprising gravitational properties of gluons would remain unknown.

Gluons are bosons, quarks and leptons, and their antiparticles are fermions. Gluons are massless (because they do not interact with the Higgs field), whereas fundamental fermions have mass because they do interact with the Higgs field. Because of these differences, which aren't the only ones, we shouldn't be surprised if there are significant differences between the gravitational properties of gluons and fundamental fermions (and antifermions). For example, if there are two equal masses, one corresponding to gluons and the other to fundamental fermions, we cannot be sure that the corresponding gravitational charges are equal.

Assuming that quarks and antiquarks have gravitational charges of opposite sign, it is obvious that the gravitational charge of sea-quarks in Table 1 is zero, while the contributions corresponding to valence quarks and valence antiquarks must be of opposite sign. If we also assume that the mass contributions of the gluons are also their gravitational charges, then (using the notations and numerical values from Table 1) we have:

$$a_{\bar{H}} = \frac{(M_{ge} + M_{ag}) - M_{vq}}{(M_{ge} + M_{ag}) + M_{vq}}\, g \Longrightarrow a_{\bar{H}} \approx 0.3g$$

This is much smaller than the preliminary ALPHA-g result $a_{\bar{H}} \approx 0.75g$, and could be an indication that the WEP is violated not only in the fermionic sector of the Standard Model of particles and fields but also for gluons. Completely terra incognita.

As a warning of how many unknowns there are, let us consider the following ad hoc relationship, which shows an intriguing numerical agreement with the ALPHA-g result.

$$a_{\bar{H}} = \frac{4(M_{ge} + M_{ag}) - M_{vq}}{4(M_{ge} + M_{ag}) + M_{vq}}\, g \Longrightarrow a_{\bar{H}} = 0.76g$$

This is probably just a coincidence, but it could also be the result of a different proportionality between mass and gravitational charge for gluons and fundamental fermions (antifermions); in the relationship above we have made the ad hoc assumption that in the gluonic sector the gravitational charge is 4 times greater than the corresponding mass.

In addition to the above question (of how to reconcile Quantum chromodynamics and a possible WEP violation), there is the question of how to reconcile the ALPHA-g result with Villata's theoretical work on General relativity and CPT symmetry [28]. In principle, a theory of gravity can be compatible or incompatible with CPT symmetry; at the present stage of our knowledge, CPT is not a criterion for the validity of a theory of gravity. Villata has shown that the compatibility of General relativity and CPT symmetry implies gravitational repulsion between matter and antimatter. How does the empirical evidence (that there is no gravitational repulsion between atoms and antiatoms) affect Villata's prediction? A possible reconciliation between the results of ALPHA-g and Villata is suggested in reference [29]. However, in our opinion there is no tension between ALPHA-g and Villata's conclusion, but it must be clearly stated that it is only valid for the fundamental particles and antiparticles of the Standard Model (quarks and antiquarks, leptons and antileptons). For composite systems (such as atoms and antiatoms) containing both the fundamental particles and the antiparticles, no general statement can be made; each composite system must be studied separately. This will become clearer in the next subsection where, in addition to antiatoms, we present the forthcoming experiments with positronium and muonium.

## 2.3. Upcoming Antimatter Gravity Experiments in the Quark and Lepton Sector of the Standard Model

### 2.3.1. Experiments with Antiatoms at CERN

ALPHA-g is just one of three competing antimatter gravity experiments at CERN, dedicated to measuring the gravitational acceleration of antiatoms in the Earth's gravitational field. Although the experiment is extremely sophisticated, the principle of the measurement is very simple. Low-energy antiprotons with a kinetic energy distribution consistent with a trap depth of about 0.5 Kelvin (temperature-equivalent energy units are used) are trapped in a vertical magnetic trap. If allowed to escape, the antiatoms will exit either at the top or at the bottom of the trap; the number of antiatoms escaping up and down can be determined by counting the annihilation events. If the magnetic trap is horizontal (and the number of antiatoms tested is large), we should expect an equal number of antiatoms to escape to the left and right. However, in the case of a vertical trap, gravity will cause a difference in the number of antiatoms escaping up and down. Comparing the number of atoms and antiatoms escaping up and down tells us whether they have the same or different (and how much different) acceleration in the Earth's gravitational field.

The next step in the ALPHA-g mission is to increase the precision of the measurements via laser-cooling of the trapped antiatoms. The current precision of about 20% will be improved to 1% in the upcoming experiments.

CERN's other two antimatter gravity experiments, The Antimatter Experiment: gravity, Interferometry, Spectroscopy (AEgIS) [30, 31] and the Gravitational Behaviour of Antihydrogen at Rest (GBAR) [32, 33], are based on very different measurement principles; AEgIS is based on interferometry, while GBAR is close to our classical vision of a free-fall experiment.

The competing AEgIS experiment will measure the vertical deviation of a pulsed horizontal beam of cold antihydrogen atoms; the vertical deviation, which is expected to be a few microns, would be measured using a Moiré deflectometer.

A horizontal beam of antiprotons enters the " Moiré " setup consisting of three equally spaced elements: two gratings and a spatially resolving emulsion detector. The two gratings with periodicity $d$ define the classical trajectories leading to a fringe pattern with the same periodicity at the position of the detector. In the transit time of the particles through the device is known, absolute force (and the corresponding acceleration) measurements are possible by employing Newton's second law of mechanics. To infer the force, the shifted position of the " Moiré" pattern must be compared with the expected pattern without force. This is achieved using light and near-field interference, the shift of which is negligible.

The third competing experiment GBAR is the closest one to our classical vision of a free-fall experiment: the measurement of the time of flight corresponding to a known change of the height.

The fixed change of the height in the GBAR experiment would be small (roughly about 20 cm) and for the measurement to be successful the initial speed of $\bar{H}$ atoms must not be bigger than a few meters per second; a speed about three to four orders of magnitude smaller than the speed of antihydrogen in the ALPHA and AEGIS experiments. Hence, while the final measurement in GBAR is more direct and simpler than in the competing experiments, the preparation of the needed ultra-cold antihydrogen is more difficult task and requires to this point the unprecedented cooling of antihydrogen. Just to get an idea about complexity common to all antimatter gravity experiments let us give a few more details. In the first step GBAR would not produce atoms of antihydrogen (antiproton and positron) but rather antihydrogen ions (antiproton with two positrons). This is motivated by the fact that ion-cooling techniques are more efficient than techniques of cooling neutral atoms. In the second step, antihydrogen ions would be sympathetically cooled with laser cooled matter ions such as $Be^+$ to temperatures of $< 10\mu K$ (i.e. with velocities of the order of 0.5 m/s). After that, the extra positron may be photo-detached by a laser pulse, with energy of only a few micro electron volt above the threshold, in order to obtain an ultracold antiatom. The time of flight of the resulting free fall should be about 200ms, which can be easily measured to extract the acceleration due to Earth's gravity.

In short, it is very likely that this decade we will see the gravitational acceleration of antiatoms measured to an accuracy of about 1% by three independent experiments using three different methods. The preliminary result of the ALPHA-g experiment suggests that even 1% accuracy could lead to the discovery of new physics.

Note that due to the complexity of creating and manipulating antiatoms, all three experiments (with the hard work of a large team of scientists) take more than two decades to produce results. And only when all 3 experiments agree will we say, yes, we know the gravitational properties of antiatoms. What a difference (unfortunately not always) between scientific attitudes and attitudes in human society.

*2.3.2. Experiments with Positronium and Muonium*

As we've seen, because of the complex structure of nucleons, experiments with anti-atoms (roughly speaking, experiments in the quark sector of the Standard Model) do not give us direct evidence about the gravitational properties of the fundamental particles of the Standard Model of particles and fields. In the lepton sector of the Standard Model, however, experiments with positronium [34, 35] and muonium [36, 37] can provide direct evidence, but these experiments are even more difficult than experiments with antiatoms because of the short lifetimes of these systems (the ground-state lifetime of positronium is about 142 ns, while that of muonium is 2.2 μs). A free-fall distance for such short lifetimes is too small to be observed. Fortunately, there are sophisticated and ingenious ideas to overcome the obstacle of short lifetimes.

Before the intriguing ALPHA-g result, antimatter-gravity experiments were regarded as research that was necessary to complete the experimental evidence, but could not produce surprising results; the mainstream thinking was that the outcome of antimatter-gravity experiments was known in advance. But now antimatter-gravity experiments, especially those with positronium and muonium, which (unlike experiments with anti-atoms) are easy to interpret, must be given high attention and priority, as perhaps top candidates for the discovery of new physics.

Positronium is a hydrogen-like atom composed of an electron and an antielectron (positron). Whatever the gravitational acceleration measured, the interpretation is unambiguous. For example, if leptons and antileptons had opposite gravitational charges, both the gravitational charge of positronium and its gravitational acceleration would be zero.

The lifetime of positronium is an increasing function of the principal quantum number $n$; intuitively it can be understood as decrease of annihilation rate because larger $n$ means a larger distance between electron and positron. For a given $n$, the lifetime is longer for the higher values of the angular momentum; the lifetime increases [35] as $n^3$ for non-circular states and $n^5$ for circular states (i.e. states with maximal angular momentum $l = |m| = n - 1$). For instance, lifetime corresponding to $n = 30$ is respectively a few milliseconds and a few seconds for non-circular and circular states.

Hence, the gravitational experiments are possible only with positronium atoms optically excited to long-lived Rydberg states (i.e., states with large $n$). The good news is that there is an encouraging initial success in creation of excited states of positronium by laser.

An experimental programme [34, 35] currently underway at University College London (UCL) has as its *long-term goal* a gravitational free-fall measurement of positronium atoms. On their long way to success, they must overcome many obstacles; among these are the production of positronium atoms in a cryogenic environment, efficient excitation of these atoms to suitably long-lived Rydberg states, and their subsequent control via the interaction of their large electric dipole moments with inhomogeneous electric fields.

Muonium is also hydrogen-like system composed from an antimuon $\mu^+$ (which is unstable with a lifetime of $2.2 \mu s$) and an electron $e^-$. It is obvious that the lifetime of the muonium is limited to the lifetime of the antimuon and cannot be made longer. Despite the obstacle of a really short lifetime (and some other obstacles), the Muonium Antimatter Gravity Experiment (MAGE) is in preparation [36, 37]. Hopefully the MAGE experiment, based on an ingenious application of well-established atomic interferometry, will be successful.

The measurement of the gravitational acceleration of muonium - if feasible - would, as the authors rightly point out, be the first gravitational measurement of a 2nd generation particle of the Standard Model. We might add that it could also be the first system to fall upwards instead of downwards. In fact, the mass of the antimuon (which is a fundamental antilepton in the Standard Model of particles and fields) is about 200 times greater than that of the electron (which is a lepton); therefore, if leptons and antileptons have opposite gravitational charges, the muon should fall upwards. ALPHA-g opened up an amusing possibility: antiatoms fall downwards, positronium doesn't fall at all, while muonium falls upwards.

In short, while CERN's experiments with antiatoms can demonstrate the existence of new physics related to the different gravitational properties of matter and antimatter, we cannot be sure of the nature of the new physics without complementary experiments with positronium and muonium.

## 3. Quantum Vacuum Cosmology

Quantum Vacuum cosmology, or QV cosmology for short, is an emerging alternative [5-10] to standard ΛCDM cosmology; QV cosmology appears to be a promising theory, but is still in its infancy. To better understand the differences between ΛCDM cosmology and QV cosmology, let us briefly recall how any general relativistic cosmology works [38].

### 3.1. How General Relativistic Cosmology Works

The cosmological principle (i.e., the assumption of the large-scale homogeneity and isotropy of the Universe) determines the Friedman-Lemaitre-Robertson-Walker (FLRW) metric:

$$ds^2 = c^2 dt^2 - R^2(t)\left[\frac{dr^2}{1 - kr^2} + r^2(d\theta^2 + sin^2\theta\, d\phi^2)\right] \qquad (1)$$

where $k = +1, k = -1 \text{ and } k = 0$ correspond respectively to a closed, open, and flat Universe, and R(t) denotes the so-called *scale factor* of the Universe. Note that, for $k = 0$, the metric is Euclidian.

Let us underscore that the real value of $k$ ($k = +1, k = -1 \text{ and } k = 0$ are reduced values) is not known and consequently the real scale factor of the Universe is also unknown. For instance, in the case of a closed Universe, the real scale factor of the Universe is its value for $k = +1$ multiplied by $\sqrt{k}$.

The dynamics of the above space-time geometry is entirely characterised by the scale factor R(t) which is the solution of the Einstein equation

$$G_{\mu\nu} = (8\pi G/c^4)T_{\mu\nu} \tag{2}$$

Einstein tensor $G_{\mu\nu}$ is determined by FLRW, but in order to solve the Einstein equation we must know the Energy-momentum tensor $T_{\mu\nu}$. Key point: Energy-momentum tensor is approximated by energy-momentum tensor of a perfect fluid characterised at each point by its proper density $\rho$ and pressure $p$.

If the cosmological fluid consists of several distinct components denoted by *n*, the results are [38] the cosmological field equations:

$$\ddot{R} = -\frac{4\pi G}{3} R \sum_n \left(\rho_n + \frac{3p_n}{c^2}\right) \tag{3}$$

$$\dot{R}^2 = \frac{8\pi G}{3} R^2 \sum_n \rho_n - kc^2 \tag{4}$$

At this point it is clear that the cosmological field equations can only be solved if we know the number of different cosmological fluids (i.e., the mass-energy content of the Universe), and how the density ($\rho_n$) and pressure ($p_n$) of each fluid evolve with the scale factor of the Universe. Here we can see the complementary nature of General relativity and the Standard Model of particles and fields (or any theory that provides a better understanding of fundamental particles and their non-gravitational interactions). Roughly speaking General relativity contributes the law of gravity and the cosmological field equations while the Standard Model contributes the content of the Universe. In simple terms, the relationship between General relativity and the Standard Model can be expressed as a question and a promise from a cosmologist to a Standard Model physicist: Please tell me the content of the Universe and I will tell you how the Universe evolves in time.

The most natural is to use the cosmological field equations with the content of the Universe given by well-established physics, i.e., the Standard Model. It is only when such theoretical predictions conflict with sophisticated astronomical observations that we need to look for new physics.

According to the Standard Model of particles and fields there are *three* different cosmological fluids, which evolve according to the power-law

$$\rho_n = \rho_{n0} \left(\frac{R_0}{R}\right)^n \tag{5}$$

where, as usually, index 0 denotes the present-day value. The corresponding equation of state, relating pressure and density is:

$$p_n = w_n \rho_n c^2 \tag{6}$$

where the equation of state-parameter $w_n = (n-3)/3$ is a constant.

More precisely these three cosmological fluids are:

Non-relativistic Standard Model matter (usually called pressureless matter or dust) with n=3 and $w_m = 0$:

$$\rho_m = \rho_{m0}\left(\frac{R_0}{R}\right)^3, p_m = 0 \qquad (7)$$

Relativistic Standard Model matter (usually called radiation) with n=4 and $w_r = \frac{1}{3}$:

$$\rho_r = \rho_{r0}\left(\frac{R_o}{R}\right)^4, p_r = \frac{1}{3}\rho_r c^2 \qquad (8)$$

Quantum vacuum with n=0 and $w_{qv} = -1$:

$$\rho_{qv} = \text{constant}, \ p_{qv} = -\rho_{qv}c^2 \qquad (9)$$

This old Standard Model description of the quantum vacuum given by equations (9), with constant density $\rho_{qv}$ and the equation of state parameter $w_{qv} = -1$, is now challenged by astronomical observations presented in Section 4, but also by explicit QFT calculations which show that this formula $w_{qv} = -1$ receives a correction [39]. As a result of these corrections, the quantum vacuum can effectively behave as quintessence or even as phantom dark energy. The departure from Eq. (9) is also inherent in the QV cosmology discussed in this paper.

Let us note that quantum vacuum (see Eq. (9)) is a fluid with negative pressure! As a simplified way to accept counterintuitive negative pressures consider the usual definition of pressure:

$$p_n = -\frac{\partial U_n}{\partial V} \qquad (10)$$

In general, if $U_n$ (the total energy of the component n) increases with the volume (i.e., with the scale factor $R$ of the Universe) the pressure is negative, if $U_n$ decreases the pressure is positive; pressure is zero only if $U_n$ is a constant. By the way, the quantum vacuum of the Standard Model of particles and fields has a constant energy density; consequently, the total energy linearly increases with the volume and the corresponding pressure is negative as in Eq. (9).

The eventual impact of a fluid with negative pressure can be understood from the cosmological equation (3). The acceleration is negative (which corresponds to the decelerated expansion of the Universe) only if the sum on the right-hand side is positive; the acceleration is positive (which corresponds to accelerated expansion of the Universe) only if the sum on the right-hand side is negative, which can be achieved with a fluid with sufficiently large negative pressure. Hence, thanks to cosmological fluids with a negative pressure General relativistic cosmology has the potential to explain both decelerated and accelerated phases in the expansion of the Universe.

Non-relativistic and relativistic Standard Model matter is such a well-established content of the Universe that it must be included in any cosmological model; hence it is included in both ☐CDM cosmology and QV cosmology.

The quantum vacuum described by Eq. (9) cannot be used as a cosmological fluid because of the cosmological constant problem (i.e. the enormous value of $\rho_{qv}$ predicted by theory). However, a fluid inspired by Eq. (9) can be used if instead of $\rho_{qv}$ we postulate a much lower density $\rho_{de}$ ($\rho_{de} \ll \rho_{qv}$) which has no theoretical background but is determined to fit observations; indeed, such a hypotheses known as dark energy is one of the ad hoc hypotheses in ΛCDM Cosmology. As we will see in subsection 3.2 the approach of QV cosmology is radically different; the quantum vacuum is a cosmological fluid and a source of gravity only because of the gravitational polarisation caused by the Standard Model matter immersed in it.

As we shall see in the different epochs of the evolution of the Universe, the effective gravitational charge density of the quantum vacuum can be a growing function, a constant function and a decaying function, and the corresponding pressure of the quantum vacuum is negative, zero and positive respectively.

In short, unlike ΛCDM cosmology which is based on a jumble of ad hoc hypotheses (in addition to dark energy there is dark matter and cosmic inflation, acting in a tiny fraction of the first second of the Big Bang event), QV cosmology is based on two postulates [7, 10]:

- *The only contents of the Universe are the quantum vacuum and the Standard Model matter immersed in it.*
- *Quantum vacuum fluctuations are virtual gravitational dipoles*

Thus, according to the first postulate, QV cosmology excludes the hypothetical dark matter and dark energy as contents of the Universe, but according to the second postulate, it considers the quantum vacuum (enriched with virtual gravitational dipoles) as the source of gravity in the Universe (a source that could eventually explain the observed phenomena without invoking contents of the Universe that are not Standard Model matter and without any modification of gravity).

If confirmed by antimatter gravity experiments, the second postulate tells us that the gravitational charge of the quantum vacuum, corresponding to virtual quark-antiquark and lepton-antilepton pairs, is zero. This is analogous to the fact that the quantum vacuum has zero electric charge because the quantum vacuum fluctuations have equal amounts of positive and negative electric charge; in fact, the electric analogy of the cosmological constant problem is prevented by the existence of virtual electric dipoles. Of course, even if quarks and leptons do not contribute to the cosmological constant problem, we still need to understand the relationship between the cosmological constant problem and gauge bosons. Therefore, the second postulate cannot be considered as a complete solution to the cosmological constant problem; however, there are both promising and attractive approaches to the cosmological constant problem and the equation of state of the quantum vacuum, such as the recent running vacuum approach [39, 40, 41], and it is an important question whether it can somehow be compatible with the existence of virtual gravitational dipoles.

Note that there is some flexibility in the above hypotheses. The Standard Model of particles and fields may be extended in the future to include additional fundamental particles, such as sterile neutrinos. It is possible that not all, but only a subset, of quantum-vacuum fluctuations are gravitational dipoles (in analogy to the fact that only a subset of quantum-vacuum fluctuations are electric dipoles), but in such a case the cosmological constant problem remains unsolved.

3.2. Prelude to QV Cosmology

The basis of the gravitational polarization of the quantum vacuum is simple and elegant [13-18].

If gravitational dipoles (with a non-zero gravitational dipole moment $\boldsymbol{p}_g$) exist, the gravitational polarization density $\boldsymbol{P}_g$, i.e. the gravitational dipole moment per unit volume, can be attributed to the quantum vacuum. It is obvious that the magnitude of the gravitational polarization density $\boldsymbol{P}_g$ satisfies the inequality $0 \leq P_g \leq P_{gmax}$, where 0 corresponds to random orientation of dipoles, while the maximal magnitude $P_{gmax}$ corresponds to the case of saturation (when all dipoles are aligned with the external field). The value $P_{gmax}$ must be a universal constant related to the gravitational properties of the quantum vacuum; a rough approximation [16] is $P_{gmax} = 3/16\pi \ (kg/m^2)$. Note that the magnitude of a single gravitational dipole in the quantum vacuum must be less than $\hbar/c$ [13, 14, 16].

If the external gravitational field is zero, quantum vacuum may be considered as a fluid of randomly oriented gravitational dipoles. In this case, the gravitational polarization density is equal to zero ($P_g \equiv 0$); of course, such a vacuum is not a source of gravitation.

However, the random orientation of virtual dipoles can be broken by the gravitational field of the immersed Standard Model matter. Massive bodies (particles, stars, planets, black holes, . . . ) but also many-body systems such as galaxies are surrounded by an invisible halo of gravitationally polarized quantum vacuum, i.e. a region of non-random orientation of virtual gravitational dipoles ($\boldsymbol{P}_g \neq 0$).

The magic of non-random orientation of dipoles, i.e. the magic of gravitational polarization of the quantum vacuum, is that the otherwise gravitationally featureless quantum vacuum becomes a source of gravity; in fact, a major source of gravity in the Universe.

Figures 1 and 2 give a mental picture of the fundamental working hypothesis of how the quantum vacuum becomes a source of gravity in the Universe.

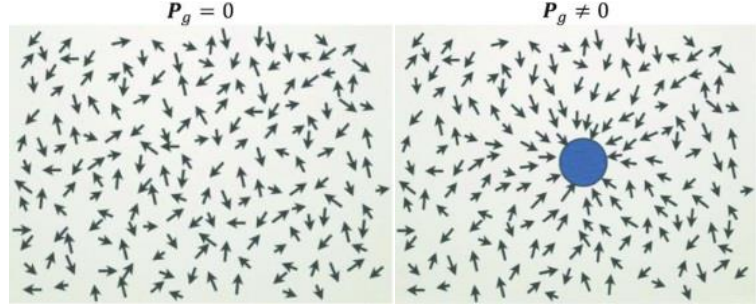

**Figure 1.** On the left, randomly oriented gravitational dipoles (in the absence of an external gravitational field); quantum vacuum is not a source of gravity. On the right, halo of non-random oriented gravitational dipoles caused by a body with baryonic mass $M_b$; the polarised quantum vacuum is a source of gravity.

### 3.3. The Fundamental Equations of the Theory

This qualitative picture presented above, can be mathematically described in the following way.

The *spatial variation* of the gravitational polarization density generates [13, 15, 16] *a gravitational bound charge density* of the quantum vacuum:

$$\rho_{qv} = -\boldsymbol{\nabla} \cdot \boldsymbol{P}_g \tag{11}$$

This gravitational bound charge density is in fact an effective gravitational charge density, which acts as if there is a real non-zero gravitational charge. Let us repeat that is how the magic of polarization works; quantum vacuum is a source of gravity, thanks to the immersed Standard Model matter.

In general, the magnitude $P_g$ is a function of the coordinates $(x_1, x_2, x_3)$, while the gravitational polarization density $\boldsymbol{P}_g$ and the corresponding Newtonian acceleration $\boldsymbol{g}_N$ (determined by distribution of the Standard model matter) point in the same direction. Consequently, we can write:

$$P_g = P_{gmax} f_g(x_1, x_2, x_3) \text{ and } \boldsymbol{P}_g = P_{gmax} f_g(x_1, x_2, x_3) \frac{\boldsymbol{g}_N}{g_N} \tag{12}$$

The values of function $f_g(x_1, x_2, x_3)$ belong to the interval [0, 1] where values 0 and 1 correspond, respectively, to the cases of random orientation of dipoles and saturation.

Equations (11) and (12) lead to

$$\rho_{qv}(x_1, x_2, x_3) = -P_{gmax} \boldsymbol{\nabla} \cdot \left[ f_g(x_1, x_2, x_3) \frac{\boldsymbol{g}_N}{g_N} \right] \tag{13}$$

Hence, in order to calculate the effective gravitational charge density of the quantum vacuum (analytically for N ≤ 2 and numerically for N ≥ 3 point-like bodies), we must know the function $f_g(x_1, x_2, x_3)$ and the unit vector $\boldsymbol{g}_N / g_N$ of the Newtonian acceleration caused by the distribution of the standard model matter.

## 3.4. Gravitational Approximation of the Quantum Vacuum by Ideal Gas of Non-Interacting Dipoles

Because of our very limited understanding of the quantum vacuum, the main question is how to find a reasonable approximation for the function $f_g(x_1, x_2, x_3)$. Unfortunately, absolutely the only option is to consider the quantum vacuum as a cosmological fluid composed of *non-interacting gravitational dipoles*, i.e. an *ideal gas* of virtual gravitational dipoles [16].

So, let us consider an ideal gas of virtual gravitational dipoles with the gravitational dipole moment $\boldsymbol{p}_g$ and energy $\varepsilon_g = -\boldsymbol{p}_g \cdot \boldsymbol{g}_N$ in a classical (Newtonian) gravitational field $\boldsymbol{g}_N$. Let us underscore that from a purely mathematical point of view, there is no difference between such a system of gravitational dipoles in an external gravitational field and well-known ideal paramagnetic gas, i.e. a system of non-interacting magnetic dipoles with the magnetic dipole moment $\boldsymbol{\mu}$ and energy $\varepsilon_\mu = -\boldsymbol{\mu} \cdot \boldsymbol{B}$ in an external magnetic field $\boldsymbol{B}$; the physical phenomena are different, but the mathematical equations are the same.

In principle an ideal gas of virtual gravitational dipoles (existing only as the quantum vacuum fluctuations) is different from an ideal gas of real (non-virtual) gravitational dipoles (which apparently do not exist).

If there were real gravitational dipoles, according to the statistical mechanics, the first step is to find the partition function, from which all macroscopic quantities (including the gravitational polarization density) can be obtained. In this case, the gravitational polarization density is a function of the well-known parameter $\beta \equiv 1/k_B T$ (where $k_B$ and $T$ denote respectively the Boltzmann constant and temperature), which is omnipresent in statistical mechanics.

However, in the case of virtual gravitational dipoles, instead of $\beta \equiv 1/k_B T$ a non-thermal parameter must be used, because, roughly speaking, quantum vacuum is a state of matter-energy at zero-temperature; to some extent like the quantum phase transitions. A quantum phase transition [42] occurs at zero temperature where thermal fluctuations are absent and instead the transition is driven by quantum fluctuations, which are tuned by variations in some non-thermal parameters; magnetic field is just one example of non-thermal parameters used instead of temperature $T$. Note that quantum phase transitions are transitions between two different states of ordinary matter immersed in the quantum vacuum; however we cannot exclude the existence of phase transitions between different states of quantum vacuum fluctuations. In principle, phase transition might exist as discontinuities of $P_g$ which, if they exist, are lost in the ideal gas approximation. However, at the present stage of theory, the limited goal is only to show (using simplified studies) that the quantum vacuum as a source of gravity (through the mechanism of the gravitational polarization) deserves more profound consideration as a plausible candidate to explain phenomena usually attributed to dark matter and dark energy.

Let us consider the simplest case of a point-like body with mass $M_b$ immersed in the quantum vacuum, with additional simplification that energy $\varepsilon_g$ can have only two different values $\varepsilon_g = \pm p_g g_N$ (in other words the angle between $\boldsymbol{p}_g$ and $\boldsymbol{g}_N$ can have only two values (0 and $\pi$)). This is an illuminating case with exact solution. Because of spherical symmetry $\boldsymbol{g}_N/g_N = -\boldsymbol{r}_0$, where $\boldsymbol{r}_0$ denotes the unit vector of the radial coordinate $r$; consequently, the fundamental equation (1) for the effective gravitational charge density reduces to:

$$\rho_{qv}(M_b, r) = \frac{1}{r^2}\frac{d}{dr}\left[r^2 P_g(M_b, r)\right] \qquad (14)$$

Just to remember, $P_g(M_b, r)$ denotes the magnitude of the gravitational polarization density $\boldsymbol{P}_g(M_b, r)$. Equation (14) leads to the following *effective gravitational charge* of the *gravitationally polarized quantum vacuum* within a sphere of radius $r$:

$$M_{gv}(M_b, r) = \int_0^r \rho_{qv}(M_b, r)dV = 4\pi r^2 P_g(M_b, r) \tag{15}$$

The corresponding acceleration caused by the quantum vacuum (which is now a source of gravity) is:

$$\boldsymbol{g}_{qv}(M_r, r) = -4\pi G P_g(M_b, r)\boldsymbol{r}_0 \tag{16}$$

Equation (16) shows that the magnitude of the gravitational acceleration caused by the quantum vacuum around a point like body is always smaller than acceleration $g_{qvmax} \equiv 4\pi G P_{gmax}$.

It is obvious that in the above equations $P_g(M_b, r) \leq P_{gmax}$ can be interpreted as the surface density of the effective gravitational charge of the quantum vacuum.

We can define a radius $R_{sat}$ so that

$$P_{gmax} = \frac{M_b}{4\pi R_{sat}^2} \Rightarrow R_{sat} = \sqrt{\frac{M_b}{4\pi P_{gmax}}} \tag{17}$$

The radius $R_{sat}$ is the radius at which the surface density corresponding to mass $M_b$ is equal to the maximum surface density $P_{gmax}$ that can be caused by the quantum vacuum around a point-like body. Roughly speaking, inside the sphere with radius $R_{sat}$ is region of saturation (in fact close to saturation). Just as an illustration, the radius $R_{sat}$ for a proton and a star as our Sun is respectively a few times $10^{-14} m$ and about 10 thousand astronomical units AU, depending on the exact value of $P_{gmax}$ which is probably [15, 16] somewhere between $0.04 \ and \ 0.06 \ kg/m^2$.

In the case of an ideal gas of virtual gravitational dipoles, the corresponding solution for the magnitude of the gravitational polarization density is

$$P_g(M_b, r) = P_{gmax} tanh\left(\frac{R_{sat}}{r}\right) \tag{18}$$

This is the first and the simplest result of the QV cosmology that should be approximately valid in the case of a point-like body, and importantly, it is a testable result. According to equations (15) and (18), the effective gravitational charge of the quantum vacuum within a sphere of radius r is:

$$M_{qv}(M_b, r) = 4\pi P_{gmax} r^2 tanh\left(\frac{R_{sat}}{r}\right), R_{sat} \equiv \sqrt{\frac{M_b}{4\pi P_{gmax}}} \tag{19}$$

The corresponding acceleration caused by the quantum vacuum surrounding a point-like body is:

$$g_{qv}(M_b, r) = 4\pi G P_{gmax} tanh\left(\frac{R_{sat}}{r}\right) < 4\pi G P_{gmax} \equiv g_{qvmax} \tag{20}$$

Note that the acceleration $g_{qv}(M_b, r)$ is always less than the apparently universal constant, estimated [16] to be $g_{qvmax} \approx 5 \times 10^{-11} \ m/s^2$. At distances $r > R_{sat}$, the quantum vacuum around a point-like body behaves like a point-like source of gravity with a gravitational field approximately proportional to $1/r$ (hence, there is a departure from the inverse square law of the immersed body). Physically there are 2 sources of gravity, a body with mass $M_b$ and the surrounding quantum vacuum. If this seemingly improbable theory is ultimately true, then the fundamental flaw of all modified gravity theories is that they erroneously assume the existence of a single physical source of gravity, when in fact there are two sources. Of course, any mathematical attempt to describe two sources of gravity as a single one inevitably looks like a violation of Newton's law of gravity. On the other hand, the Dark Matter paradigm is correct in that there are two sources of gravity, but without the understanding that the second source is the quantum vacuum, and not some as-yet unknown kind of matter completely outside the Standard Model of particles and fields.

At this point, let us make an interlude to show that the first and the simplest result of the theory, i.e. Eq. (18) and its consequence Eq. (20), is already an astonishing success, allowing a quantitative explanation by the quantum vacuum of the galactic phenomena usually attributed to dark matter or the modification of gravity [16, 43].

*3.4.1. Interlude – Dark Matter the First Quantitative Success of QV Cosmology*

The total acceleration $g_{tot}(M_b, r)$ caused by a point like body of mass $M_b$ at distance $r$ is a sum of the Newtonian acceleration $g_N(M_b, r)$ and acceleration $g_{qv}(M_b, r)$ caused by the quantum vacuum.

$$g_{tot}(M_b, r) = g_N(M_b, r) + g_{qv}(M_b, r) = g_N(M_b, r) + 4\pi G P_{gmax} tanh\left(\frac{R_{sat}}{r}\right) \qquad (21)$$

So the acceleration is greater than Newtonian, but the extra acceleration is not caused by dark matter or a modification of gravity (as in the Modified Newtonian Dynamics (MOND)). The extra acceleration $g_{qv}(M_b, r)$ is caused by the gravitational polarisation of the quantum vacuum, by the immersed Standard Model matter. Note that the gravitational field of the quantum vacuum can be neglected, i.e. the total acceleration $g_{tot}(M_b, r)$ is well approximated by the Newtonian acceleration $g_N(M_b, r)$ if the Newtonian acceleration is much larger than $4\pi G P_{gmax}$, which is the maximum of the magnitude of the acceleration that can be caused by the quantum vacuum; within the framework of QV cosmology the existence of this maximum is the reason for the success of the otherwise physically incorrect MOND theory.

The striking surprise is that, in the case of a galaxy, this additional acceleration, described by the seemingly oversimplified equation (18), leads to results that agree well with predictions based on dark matter and MOND [16, 43]. Let us explain this in more detail.

MOND is an ad hoc theory that is very successful for individual galaxies. Even the strongest proponents of the dark matter paradigm admit the 'unreasonable effectiveness' of MOND at the galactic scale; with the current precision of galactic measurements we cannot distinguish between these two theories. Thus, at the galactic scale, agreement with one of these two theories means agreement with the other; we explain here the agreement already established with MOND.

The starting point of MOND is an ad hoc assumption that, for a point-like source of gravity, the ratio of total to Newtonian acceleration ($g_{tot}/g_N$) is a function of the ratio $a_0/g_N$ of a universal acceleration $a_0 \approx 1.2 \times 10^{-10} ms^{-2}$ and the Newtonian acceleration, i.e. $g_{tot}/g_N = f(a_0/g_N)$. Various (ad-hoc) interpolating functions $f(a_0/g_N)$ are used to *fit galaxy rotation curves*; the most successful interpolating functions are the simple, standard and radial acceleration relation (RAR). The key point is that (as shown in Section 6 of reference [16] and confirmed by eminent MOND researchers in reference [43]) the fundamental equation (18) determines an interpolation function (Eq. (33) in reference [16]) which has numerical values very close to the values of the interpolation functions used in MOND (in fact the graph of the QV interpolating function lies mainly between the graphs of the two MOND functions; see Appendix B.2).

In conclusion, the interpolation function "coming" from the quantum vacuum, with a clear physical hypothesis behind it, is just as valuable (i.e. indistinguishable from others with the current accuracy of astronomical measurements) as the ad hoc (and therefore unphysical) functions used in MOND. Since all MOND results follow from the interpolating function, and QV cosmology provides such a function (indistinguishable from others at present), this means that the gravitational polarisation of the quantum vacuum leads to a quantitative description of the galaxy that is as good as MOND's description [16, 43]

Note that reference [43] acknowledges the value of the fundamental equation (18) and its consequences on a galactic scale, but claims that it is in conflict with existing ephemerides on a solar system scale. However, their conclusion about the validity of equation (18) within the solar system is based on plausibility arguments rather than a rigorous procedure; this is questionable as explained and warned in section 5.3 of the excellent review "Testing theories of gravity with planetary ephemerides" [44]. A typical example of plausibility arguments is: the tiny contribution of the quantum vacuum to the precession of a planet's orbit is greater than the uncertainty in the perihelion advance per orbit, and so the effect of the quantum vacuum is dismissed. This seems reasonable at first sight. However, the rigorous and much more complex approach is to compare the ephemerides corresponding to the two dynamical models [44]. Just to give you an idea of how large the contribution of the gravitational polarisation of the quantum vacuum is, note that the total effective gravitational charge of the quantum vacuum within the Earth's orbit is equal to $8.4 \times 10^{-9} M_\odot$, where $M_\odot$ is the mass of the Sun. In short, while the validity of Eq. (21) on a very small scale such as the Solar System is an open question, it is an astonishing success on a galactic scale.

It is very important to emphasise that within the ΛCDM cosmology, dark matter complements Standard Model matter in two crucial ways. First, the observed dynamics of existing galaxies is a joint effect of Standard Model matter and dark matter. In addition to this role in ensuring the longevity of galaxies, dark matter also ensures the formation of galaxies in the Universe. In fact (assuming that our law of gravity is correct and that there is no additional source of gravity), the gravitational field created by the initial hydrogen and helium gas is not strong enough to create galaxies. QV cosmology must therefore explain how the quantum vacuum and Standard Model matter manage to create the potential wells necessary for galaxy formation; such a possibility is suggested at the end of the next subsection.

*3.4.2. A Generalisation of the Solution Based on the Ideal Gas Approximation*

Let us emphasise the major shortcoming of the ideal gas approximation for $r \gg R_{sat}$. According to Eq. (19), the effective gravitational charge of the quantum vacuum in a sphere around a body is a linear function of the radius of the sphere, leading to the erroneous conclusion that the halo of the polarised quantum vacuum can be infinitely large. In fact, each halo has a maximum possible size [8] roughly determined by a characteristic radius $R_{ran}$ ($R_{ran} \gg R_{sat}$), because at distances greater than $R_{ran}$, the gravitational field is not strong enough to cause gravitational polarisation; the dipoles are randomly oriented and the effective gravitational charge density outside the sphere of radius $R_{ran}$ is zero everywhere. Adding this feature, which is lost in the ideal gas approximation, the general form of Eq.(19) can be written as:

$$M_{qv}(M_b, r) = 4\pi P_{gmax} r^2 tanh\left(\frac{R_{sat}}{r}\right) f(R_{ran}, r) \qquad (22)$$

It is obvious that the unknown function $f(R_{ran}, r)$ must satisfy two conditions. Firstly, in order to maintain equation (18) in the region of its validity, it is necessary to have $f(R_{ran}, r \ll R_{ran}) = 1$. Second, in order to obtain a constant value in the region of random orientation, the function $f(R_{ran}, r \gg R_{ran})$ must be inversely proportional to $r$. A simple interpolating function that satisfies these conditions is

$$f(R_{ran}, r) = tanh\left(\frac{R_{ran}}{r}\right) \qquad (23)$$

Of course, the choice of the interpolating function $f(R_{ran}, r)$ is not unique; the function (23) is only a rough working approximation (or toy model) that is expected to give the correct qualitative and rough quantitative behaviour. Note that an analogous situation occurs in many emerging theories. For example, there are different interpolating functions in MOND, different empirical laws for the distribution of dark matter, and many different functions for the inflation field in cosmic inflation theory.

It is obvious that an observer at a distance $r$ from the point-like body measures the mass $M_b$ (i.e. the gravitational charge) of the body plus the effective gravitational charge $M_{qv}(M_b, r)$ of the quantum vacuum within the corresponding sphere of radius $r$, The total gravitational charge

$$M_{tot}(M_b, r) = M_b + M_{qv}(M_b, r) = M_b + 4\pi P_{gmax} r^2 tanh\left(\frac{R_{sat}}{r}\right) tanh\left(\frac{R_{ran}}{r}\right) \quad (24)$$

The main prediction of equation (24) is that two observers at different distances $r_1$ and $r_2$ will measure different central gravitational charges, i.e. $r_2 > r_1 \Rightarrow M_{tot}(M_b, r_2) > M_{tot}(M_b, r_1)$. In general the function $M_{tot}(M_b, r)$ increases from its minimum $M_b$ to its horizontal asymptote $M_b + M_{qvmax}(M_b)$. It is obvious that $M_{qvmax}(M_b) \equiv 4\pi P_{gmax} R_{sat} R_{ran}$ denotes the maximum possible effective gravitational charge that a body of mass $M_b$ can have, or in other words the effective gravitational charge of the halo of the maximum size.

Note that, for any interpolating function, $R_{ran}$ can be expressed as a function of $R_{sat}$. In fact, the asymptotic behaviour $M_{qv}(M_b, r \gg R_{ran}) = M_{qvmax}(M_b)$, leads to the proportionality:

$$R_{ran} = \frac{M_{qvmax}(M_b)}{M_b} R_{sat} \quad (25)$$

Let us look at Figures 2 and 3, which provide a useful mental picture and help to memorise the main effects of the gravitational polarisation of the quantum vacuum caused by a point-like body.

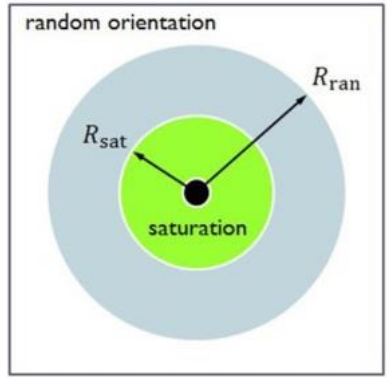

**Figure 2**. Schematic presentation of the halo regions of the polarized quantum vacuum around a point-like body: the region of saturation (green region inside the sphere with radius $R_{sat}$), the region of partial alignment of gravitational dipoles (blue) and the region of random orientation of dipoles (white region outside the sphere with radius $R_{ran}$).

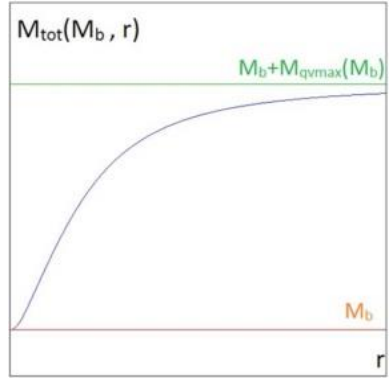

**Figure 3.** Schematic presentation of total gravitationa charge $M_{tot}(M_b, r)$ within a sphere of radius $r$. The red line is total mass (equal to $M_b$) with the neglected quantum vacuum. The blue line is the gravitational charge with the included quantum vacuum, which tends asymptotically to the constant value $M_b + M_{qvmax}(M_b)$.

From the point of view of eq. (24) and eq. (25), consider a rarefied, nearly homogeneous and isotropic gas with a very large number of N pointlike bodies. If the distances between the bodies are sufficiently large, then according to eq. (24) each body should have its own individual halo of polarised quantum vacuum of maximum size around it. (For more details, see [16] and Chapter 3 of reference [18], which also contains an equation of state for this simple cosmological fluid). The state of the Universe before the formation of structures was an analogous rarefied gas of hydrogen and helium, with each atom having enough space for the halo of maximum size and thus interacting gravitationally as if its gravitational charge were about $R_{ran}/R_{sat}$ times greater than its inertial mass $M_b$. It is to be hoped that a study of structure formation under these conditions will be carried out soon, but it seems quite plausible that initial much stronger gravitational interactions between atoms (due to the gravitational polarisation of the quantum vacuum) may be sufficient to initiate the process of structure formation.

*3.4.3. The Universe before the Formation of the First Stars and Galaxies*

If Standard Model matter and quantum vacuum are the only contents of the Universe, then before the formation of the first stars and galaxies, the Universe was, roughly speaking, just an expanding gas with only tiny deviations from perfect homogeneity and isotropy. After the birth of the Cosmic Microwave Background, it was an atomic gas composed of hydrogen and helium (with traces of deuterium and lithium) and mixed with photons and neutrinos; before the CMB, it was just plasma from which atomic gas emerged.

Different bodies compete for the gravitational polarisation of the quantum vacuum. For example, because of its proximity to the Sun, the Earth's gravitational field dominates only a small region around the Earth, and so the Earth's halo of polarised quantum vacuum is very small; if the Earth were at the edge of the Solar System, its area of gravitational dominance, and hence its halo, would be much larger. Eventually, when all other bodies are sufficiently distant, the Earth can form a halo of maximum size.

In principle, each atomic nucleus can have its own halo of polarised quantum vacuum, but just as the Earth's halo is suppressed by the gravitational field of the Sun, individual atomic halos are suppressed by the much stronger gravitational fields of stars and galaxies. However, individual atomic halos were possible when the Universe was an expanding gas.

In a sufficiently dense gas, individual halos of nuclei do not exist, and the gravitational charge of a nucleus (or atom) is equal to its mass. In a sufficiently rarefied and nearly homogeneous and isotropic gas, all nuclei can have their own halo of maximum size; since each halo has an effective gravitational charge, a nucleus or atom has a gravitational charge much greater than its mass. Consequently, the expansion of the cosmic gas leads to an increase in the effective gravitational charge of the quantum vacuum in the Universe, but the increase in the effective gravitational charge remains a constant once all atoms have reached a halo of maximum size. See Appendix B, in particular Appendix B.3, for analytical support of these and subsequent qualitative arguments.

During the increase of the effective gravitational charge, the polarised quantum vacuum acts as a cosmological fluid with negative pressure, causing the accelerated expansion of the universe. In the subsequent epoch of constant gravitational charge, the quantum vacuum acts as a pressureless cosmological fluid (Chapter 4 in [18] and Appendix B).

Imagine a *perfectly* homogeneous and isotropic gas in which (due to the absence of inhomogeneity) the gravitational formation of stars and galaxies is impossible. Due to the very large effective gravitational charge of the quantum vacuum, such a "gas Universe" would be forced to collapse, leading to a "big crunch" and a new cycle of the Universe. Note that during the collapse, after a period of constant value, there will be a period of decrease in the effective gravitational charge of the quantum vacuum in the Universe; this means that the quantum vacuum will act as a cosmological fluid with positive pressure.

Apparently (we will discuss this later) such a collapse of the "gas universe" is prevented by tiny deviations from homogeneity and isotropy, which lead to the formation of stars and galaxies.

At the birth of the CMB (roughly at redshift $z = 1100$), the average distance between nuclei was about $1.5 \times 10^{-3} m$; about ten orders of magnitude larger than saturation radius $R_{sat}$ of helium. Thus, it seems plausible that prior to the birth of the CMB, all individual halos had maximum size and maximum effective gravitational charge. This epoch was preceded by an epoch of halo growth and corresponding growth of the effective gravitational charge of the quantum vacuum, i.e. a period of early accelerated expansion of the Universe when it was in the state of plasma.

The challenging crucial question is if the gravitational polarisation of the quantum vacuum is compatible with the existing empirical evidence related to the Cosmic Microwave Background.

*CMB and gravitational polarization of the quantum vacuum*

The inevitable common point between ΛCDM cosmology and any other cosmology (including QV cosmology) is that before the birth of the CMB, the Universe contains baryon-photon fluid composed of baryons and photons interacting through radiation pressure. The difference is that in ΛCDM cosmology the baryon-photon fluid is complemented by dark matter, whereas in QV cosmology there is no dark matter. Instead of dark matter, as a consequence of the low density of baryons and the maximum size of their halos of the polarised quantum vacuum, each baryon has an effective gravitational charge $m_g$ that is $\alpha_g$ times larger than its inertial mass $m$ (i.e. $m_g = \alpha_g m$). The question is whether there is a value $\alpha_g$ for which the baryon-photon fluid without dark matter is compatible with the CMB power spectrum. Note that if $\alpha_g$ is found, it is a good approximation for the fundamental ratio $M_{qvmax}(M_b)/M_b$. Existing studies of what would be the CMB power spectrum without dark matter do not agree satisfactorily with the empirical evidence, but $\alpha_g^2$ times stronger gravitational interactions between baryons may lead to an agreement.

*3.4.4. Universe with Stars and Galaxies*

The starting point for the formation of the first stars is an expanding rarefied hydrogen-helium gas, each atom of which has an effective gravitational charge $\alpha_g$ times its mass. Imagine the formation of a star from an extremely large number $N$ of atoms; for example, mass of our Sun is equal to mass of about $N = 3 \times 10^{58}$ atoms of hydrogen.

From the point of view of the gravitational polarisation of the quantum vacuum, the formation of a star is the transformation of $N$ individual atomic halos of the polarised quantum vacuum into a single collective halo of the star; thus individual halos are lost but the halo of the star appears.

However, the volume of the maximum halo of the created star is $\sqrt{N}$ times larger than the volume of $N$ individual maximum size halos. Thus, in principle, a Universe that is large enough to accommodate all the individual maximum size atomic halos may be too small to accommodate all the macroscopic maximum size halos. Consequently, if there is a merging of maximum size atomic halos into a non-maximum size macroscopic halo, there will be a decrease in the effective gravitational charge of the quantum vacuum. QV cosmology predicts that in the epoch of structure formation, the quantum vacuum acts as a positive pressure cosmological fluid; as the universe expands, the non-maximum size halos of structures grow to their maximum size, and the quantum vacuum changes from a positive pressure cosmological fluid to a negative pressure fluid, producing the effects we now (perhaps erroneously) attribute to cosmological constant type dark energy.

Although the expansion of a Universe full of galaxies and galaxy clusters is much more complex than the expansion of a "gas Universe", the result is analogous; as the Universe expands, the quantum vacuum becomes a pressureless cosmological fluid. Due to the effective gravitational charge of the quantum vacuum which is again $\alpha_g$ times greater than the inertial mass of the Standard Model matter, the expansion of the Universe would end and the contraction would begin. Because of the gravitational version of the Schwinger mechanism, the possible end of the collapse is surprising.

*3.4.5. A Cyclic Universe with Cycles Alternately Dominated by Matter and Antimatter*

Our Universe is dominated by matter despite theoretical prediction and experimental evidence that matter and antimatter are always created in the same quantities. The Big Bang (or something similar, that we cannot still distinguish from the Big Bang) should have created equal amounts of matter and antimatter in the early Universe; one of the greatest challenges is to figure out what happened to the antimatter.

The mainstream speculation is that the matter-antimatter asymmetry is caused by CP violation, but don't be misled by the known CP violation. The known CP violation cannot produce the excess matter for a single galaxy, and there are hundreds of billions of galaxies in the visible universe. So it must be an unknown (and therefore hypothetical) type of CP violation, many orders of magnitude stronger than the known one!

According to QV cosmology, we may be living in a cyclic Universe, with cycles alternately dominated by matter and antimatter [14, 15]. This provides an elegant explanation for the matter-antimatter asymmetry in the Universe: our cycle of the Universe is dominated by matter because the previous cycle was dominated by antimatter (and the next cycle would be dominated by antimatter again).

This cyclic scenario is possible because at a sufficiently small (but still macroscopic) scale factor $R(t)$ of the Universe, the gravitational field can become strong enough to create a huge number of particle-antiparticle pairs from the physical quantum vacuum, with the additional feature that matter tends to approach the eventual singularity while antimatter is violently ejected farther and farther from the singularity. The amount of antimatter created is equal to the decrease in the mass of the collapsing matter Universe. Thus the amount of matter decreases while the amount of antimatter increases by the same amount; the result could be the conversion of almost all matter into antimatter. If the conversion process is extremely fast (and calculations show that it can be), it may look like a Big Bang, but it is not a Big Bang: it starts from a macroscopic initial size without a singularity and without the need for an inflationary field of an unknown nature.

*3.4.6. Is Our Universe One of Many Universes?*

A number of respected physical theories suggest that we live in a Multiverse, that our universe is just one of many universes. For example, there are theories of the inflationary multiverse, the brane-world multiverse, the holographic multiverse, the loop-quantum multiverse... All of these theories lead to very different models of a Multiverse.

There is an amusing possibility of the multiverse related to QV cosmology.

The metrics used in contemporary cosmology (Eq. (1)) for a closed Universe (k=1) is mathematically identical to the metrics of a closed three-dimensional volume (3-sphere) within a four-dimensional Euclidean space. Note that the physical existence or non-existence of such a four-dimensional Euclidean space has no effect on the intrinsic geometry we use.

We cannot exclude the possibility that a four-dimensional Euclidean space does exist, and that our Universe is just a three-dimensional Riemannian subspace within it. Assuming that four-dimensional Euclidean space is the physical reality, it would be strange to have only a single three-dimensional Riemannian subspace of such complexity as our Universe within it. It seems more plausible that four-dimensional Euclidean space contains an infinite number of universes (i.e. three-dimensional Riemannian subspaces). But according to General relativity, a four-dimensional space containing the enormous gravitational charges of all universes cannot be Euclidean; the only way it can be Euclidean is if the total gravitational charge of all universes is zero, and this is only possible if there are an equal number of universes with positive and negative gravitational charges.

We have seen that, according to QV cosmology, our Universe can be a cyclic one, with cycles alternately dominated by matter and antimatter, as should be the case in other universes. It is therefore natural to expect a multiverse with equal amounts of matter and antimatter, and with matter and antimatter universes equally distributed. What a nice possibility; matter-antimatter symmetry is violated in each of the individual universes, but is restored in the multiverse (which is four-dimensional Euclidean space).

Of course, the purpose of these wild speculations is not to claim that this is so, but to demonstrate once again how rich the consequences can be of a single working hypothesis that quantum-vacuum fluctuations are virtual gravitational dipoles. There is a significant possibility that QV cosmology is wrong, but it is intriguing that whatever the problem, from dark matter and dark energy to the multiverse, it has astonishing predictive power.

### 3.4.7. The Ideal Gas Approximation for $N \geq 2$ Bodies

The ratio $R_{sat}/r$ in equation (18) has meaning only in the case of a point-like body, but can be easily transformed to the following form that can be used for $N \geq 2$ bodies in any gravitational field with the Newtonian magnitude $g_N$

$$\frac{R_{sat}}{r} = \frac{1}{r}\sqrt{\frac{M_b}{4\pi P_{gmax}}} \equiv \sqrt{\frac{g_N}{4\pi G P_{gmax}}} \equiv \sqrt{\frac{g_N}{g_{qvmax}}} \qquad (26)$$

Consequently Eq.(18) which is valid only for a point-like body can be given more general form that is valid for any gravitational field $\boldsymbol{g}_N$.

$$\boldsymbol{P}_g = P_{gmax} tanh\left(\sqrt{\frac{g_N}{4\pi G P_{gmax}}}\right)\frac{\boldsymbol{g}_N}{g_N} \qquad (27)$$

According to equations (12) and (27) the unknown function $f_g(x_1, x_2, x_3)$ is equal to $tanh(\sqrt{g_N/4\pi G P_{gmax}})$.

The above generalisation (Eq.(25)) has allowed the exact solution in the case of two point-like bodies immersed in the quantum vacuum [42].

So far we considered the simplest case when energy $\varepsilon_g = -\boldsymbol{p}_g \cdot \boldsymbol{g}_N$ can have only two different values ($\varepsilon_g = \pm p_g g_N$). It is easy to generalize to the case when (as in the analogous case of paramagnetism) there are $2J + 1$ different energy levels (where J can be a positive half integer or integer number). In this more general case, hyperbolic tangent in above equations must be replaced by the corresponding Brillouin function. In the limit when $J$ tends to infinity, the Brillouin function $B_J(x)$ reduces to Langevin function $L(x)$ which describes a classical system with continuous spectrum of energy which can have any values in the interval $[-p_g g_N, p_g g_N]$. In general for a given $x$ and $J \geq 1$, $tanh(x) > B_J(x) > L(x)$; hence numerical values obtained using hyperbolic tangent and Langevin function can be respectively considered as an upper and lower bound.

If the angle between $\boldsymbol{p}_g$ and $\boldsymbol{g}_N$ can have 3 values $(0, \pi/2, \pi)$; note that this is analogous to 3 states of particle with spin one, instead of equation (27) the result is

$$\boldsymbol{P}_g = P_{gmax} \frac{2\sinh\left(\sqrt{\frac{g_N}{4\pi G P_{gmax}}}\right)}{1 + 2\cosh\left(\sqrt{\frac{g_N}{4\pi G P_{gmax}}}\right)} \frac{\boldsymbol{g}_N}{g_N} \equiv P_{gmax} B_1\left(\sqrt{\frac{g_N}{4\pi G P_{gmax}}}\right) \frac{\boldsymbol{g}_N}{g_N} \qquad (28)$$

Alternatively this can be written as

$$\boldsymbol{P}_g = P_{gmax} \frac{2\cosh\left(\sqrt{\frac{g_N}{4\pi G P_{gmax}}}\right)}{1 + 2\cosh\left(\sqrt{\frac{g_N}{4\pi G P_{gmax}}}\right)} \tanh\left(\sqrt{\frac{g_N}{4\pi G P_{gmax}}}\right) \frac{\boldsymbol{g}_N}{g_N} \qquad (29)$$

showing that the magnitude of the gravitational polarisation density corresponding to Eq. (27) is a little bit larger than the magnitude given by Eq. (29). As the difference is not very large, from the mathematical point of view (especially at this stage of toy models) it is better to use the simpler equation (27), while from a physical point of view equation (29) should be no less important.

## 4. Astronomical Challenges to the Standard ΛCDM Cosmology

A few weeks away from the James Webb Space Telescope's third birthday, it is perhaps no exaggeration to say that it is the most successful 3-year-old ever in the history of fundamental research; a 3-year-old that is already challenging the standard ΛCDM cosmology (e.g. [46-51] and references therein). In addition there are challenges coming from other instruments such as DESI (Dark Energy Spectroscopic Instrument) [52]. Let us mention three major challenges as an illustration of differences between ΛCDM cosmology and QV cosmology.

### 4.1. Evolving Dark Energy

In ΛCDM Cosmology, dark energy behaves as a cosmological constant, inspired by Eq. (9); the density of dark energy in the universe remains constant despite expansion. That's why recent evidence from the DESI (Dark Energy Spectroscopic Instrument) that the density of dark energy may be evolving (more precisely, dark energy may be weakening) is a potentially major challenge to ΛCDM Cosmology (e.g. [52] and references therein).

If the evolving dark energy is confirmed, it will be very bad news for ΛCDM cosmology; conversely, QV cosmology must be rejected immediately if it is confirmed that what we call dark energy behaves as a cosmological constant.

According to QV cosmology (as explained in Section 3.4), in an expanding universe full of galaxies, the quantum vacuum acts as a cosmological fluid with a negative pressure that decreases to zero with expansion ([15, 17] and Chapter 4 in the preprint of the book [18]); such a global behaviour of the polarised quantum vacuum is radically different from the paradigm of the cosmological constant.

## 4.2. Accelerated Formation of Galaxies

Observations from the James Webb Space Telescope suggest an accelerated formation of the first stars and galaxies ([47, 48] and references therein). Apparently, stars and galaxies are forming much faster than is possible in ☐CDM cosmology; we are observing galaxies that shouldn't exist. Bright and massive galaxies existed as early as 300 Myr after the Big Bang [47], but there is still unconfirmed evidence of galaxies as young as 200 million years after the Big Bang.

One of the most important results is the discovery and study [38] of three early ultra-massive galaxies ($log\, M_*/M_\odot \gtrsim 10^{11}$). The results, based on the confident spectroscopic redshift and stellar mass measurements, provide strong evidence that the early Universe must have been two to three times more efficient in forming massive galaxies than the average trend found by previous studies at later times. An extraordinary 50% of baryons are converted into stars - two to three times higher than even the most efficient galaxies at later epochs.

Before the first stars and galaxies formed, according to both the Standard Model of particles and fields and General relativity (and supported by the study of the Cosmic Microwave Background), the Universe was an expanding gas of atoms of hydrogen and helium (with traces of lithium), photons and neutrinos; the mass of hydrogen in the Universe was about three times that of helium, i.e. the ratio of hydrogen atoms to helium atoms was about 12:1.

The fundamental question is how the first stars and galaxies were formed from this initial atomic gas. The surprising answer is that this gas is insufficient to form stars and galaxies because there is too little baryonic mass in the hydrogen-helium gas and the associated gravitational field is not strong enough to ensure structure formation in the Universe. A possible solution proposed by ΛCDM cosmology is to add hypothetical dark matter as an extra component to the Universe, which would increase the overall density of matter and the strength of the gravitational field.

In Quantum Vacuum Cosmology there is no dark matter; a rarefied hydrogen-helium gas, photons and neutrinos are the only contents of the Universe at that time. Instead of dark matter, *each atom has an effective gravitational charge* several tens of times greater than the atom's mass; while the effective gravitational charge of such a baryonic gas in Quantum Vacuum Cosmology is several times greater than the total mass of baryonic gas and dark matter in ΛCDM Cosmology. Such an effective gravitational charge of each atom is caused by the gravitational polarization of the quantum vacuum.

In fact, as noted above, immediately after the birth of the Cosmic Microwave Background, the average distance between atoms in the nearly homogeneous and isotropic hydrogen-helium gas was about $1.5 \times 10^{-3} m$, and perhaps each atom had its largest possible individual halo of the gravitationally polarised quantum vacuum. Atomic halos of the maximum size and maximum effective gravitational charge are even more plausible at ten times smaller redshift ($z = 110$), where the Universe is still expanding hydrogen-helium gas with an average distance between atoms of about $1.5 cm$.

Although the value of the number $\alpha_g$ introduced in Section 3.4.2 is not known, we can make a seemingly reasonable estimate of its lower bound. Indeed, the estimated effective gravitational charge of galaxies is typically an order of magnitude greater than the baryonic mass and it is likely that the lower limit of $\alpha_g$ is significantly greater than 10. Knowing that mass of hypothetical dark matter in the Universe is about 6 times larger than baryonic mass, the effective gravitational charge of baryons before creation of the first star and galaxies should be $\alpha_g/6$ times larger than gravitational charge of dark matter. Hence, in QV cosmology formation of stars and galaxies must be much faster than in ΛCDM Cosmology.

### 4.3. The Hubble Tension

The Hubble constant $H_0$ (i.e. the present day value of the Hubble parameter $H$) is one of the most important numbers in cosmology because it tells us how fast the Universe is expanding. The Hubble constant can be obtained *directly* from measurements of the distances and redshifts of nearby galaxies; the result $H_0 = 73.0 \pm 1.0 \, km \, s^{-1} Mpc^{-1}$ is independent of cosmological models. Instead of a direct measurement, the Hubble constant can be derived using the ΛCDM model, with parameters calibrated to fit the CMB data; the result is $H_0 = 67.4 \pm 0.5 km \, s^{-1} Mpc^{-1}$. The Hubble tension refers to this difference between direct measurements of $H_0$ and indirect measurements given a cosmological model. Hence, for some unknown reason, the ΛCDM cosmological model doesn't produce the measured value (see [49] and references therein). If the direct measurements are accurate, then the Hubble tension is evidence for new physics (for a review, see [50]).

One of the proposals to resolve the Hubble tension is the early dark energy [51]. It is a modification of ΛCDM cosmology by introducing a new component into the energy density of the Universe; the new component behaves like dark energy in the period between matter-radiation equality (about 50,000 years after the Big Bang) and the birth of the CMB (about 380,000 years after the Big Bang). Let us recall that matter-radiation equality refers to the point in the early universe when the density of matter became equal to the density of radiation (photons and relativistic particles). Matter-radiation equality is therefore the transition point from a radiation-dominated universe to a matter-dominated universe. The appearance and disappearance of such a dark energy-like component is, of course, an ad hoc hypothesis.

Note the coincidence; in order to resolve the Hubble tension, the invoked early dark energy must exist in the same period where, according to QV cosmology (see Appendix B.3 and Section 3.4.1), the quantum vacuum acts as a cosmological fluid with negative pressure.

## 5. Different Possibilities for the Existence of Gravitational Dipoles

It is useful to recall that in electrodynamics an electric dipole is composed of a positive and a negative electric charge (e.g. virtual electron-positron or quark-antiquark pairs in the quantum vacuum!); the essence is that there are positive and negative electric charges that exist as electric monopoles. In contrast, a magnetic dipole does not consist of magnetic charges (magnetic monopoles), but is a result of the dynamics of electric charges. Therefore, dipoles can be realised in fundamentally different ways and can exist with or without the presence of the corresponding monopoles.

Here we give an overview of different possibilities for the existence of gravitational dipoles; hopefully other theoretical possibilities will be revealed in the near future. First, we point out what would be the best, i.e. an ideal kind of virtual gravitational dipole from the point of view of the gravitational polarisation of the quantum vacuum; it is a dipole that interacts only through gravity. Since there are no candidates for an ideal dipole in the Standard Model of particles and fields, we then focus on possibilities within the Standard Model. We then consider possibilities for gravitational dipoles within the framework of general relativity and its extensions.

### 5.1. Sterile Neutrino – A Candidate for the Best Kind of Virtual Gravitational Dipoles

From the point of view of the gravitational polarization of the quantum vacuum, it is obvious that *the best kind of gravitational dipole is a dipole which interacts only through gravity*. Other interactions can prevent the gravitational polarization; for instance, if a gravitational dipole is also an electric dipole, because of the fact that gravitational force is many orders of magnitude weaker than electromagnetic force, polarization is prevented.

Within the Standard Model of Particles and Fields there are no candidates for this best kind of gravitational dipole, simply because the Standard Model has no particles which participate *only* in gravitational interactions. However, there is strong theoretical and experimental motivation to complete the Standard Model of Particles and Fields with additional neutrinos, so called sterile neutrinos (for a recent review see [53], which are expected to interact only through gravity. It seems likely that sterile neutrinos exist, and if it is confirmed by experiments, we will have a candidate for the best kind of virtual gravitational dipoles.

5.2. Can Gluons Be Virtual Gravitational Dipoles?

In the absence of ideal candidates which interact only through gravity, the remaining good candidates within the current Standard Model of Particles and Fields, are only those constituents of the Standard Model which have no electric charge; hence, already known neutrinos, photons, gluons, $Z^0$ boson and the Higgs Boson.

Could gluons somehow be a surprise, i.e. virtual gravitational dipoles in the quantum vacuum? According to our current understanding of gravity, this is extremely unlikely. But if it is eventually established that lepton-antilepton and quark-antiquark pairs are gravitational dipoles, then everything will have to be reconsidered and doubted.

Strong interactions described by Quantum chromodynamics and gravitational interactions described by General Relativity (which is not a quantum theory) look so different that any physical relation between them seems very unlikely; but who knows. As an encouragement to the open-minded attitude, let us point to a recently discovered correspondence (named *the double copy*) between scattering amplitudes (quantities related to the probability for particles to interact) in gravity and Quantum chromodynamics. For instance, the double copy expresses graviton tree amplitudes (i.e. a tree-level Feynman diagrams) in terms of sums of products of gluon tree amplitudes. No one knows if these mysterious connections between gauge and gravity theories (in particular Quantum chromodynamycs and gravity) have a physical meaning, or it is just a formal, mathematical similarity between otherwise completely unrelated physical systems. In any case (see for instance two recent reviews [54, 55]) it is intriguing that, within the double-copy framework for gravity, gravity directly follows from Quantum chromodynamics, as stated already in the amusing title The double copy: gravity from gluons" [54].

If there is some physics behind the double copy correspondence, even firmly closed questions can be reopened. For example, we all (including the author of this review) believe that there is irrefutable evidence that SU(3), and not U(3) which contains SU(3) as its subgroup, is the local gauge symmetry of Quantum chromodynamics. At first sight, the U(3) group is a very plausible candidate for the symmetry of Quantum chromodynamics. In fact, SU(3) is a subgroup of U(3), but in addition to the eight generators of SU(3), and hence octet of eight gauge bosons called gluons, the U(4) has an additional U(1) generator corresponding to a ninth gauge boson, which we can call the "ninth gluon". This "ninth gluon" acts as the gauge boson of an infinite-range force [56, 57]. If the "ninth gluon" is photon it would be very beautiful unification of strong and electromagnetic interactions; if the "ninth gluon" is graviton it would be very beautiful unification of strong and gravitational interactions. The "ninth gluon" cannot be a photon [56], because the corresponding interaction looks very much like an extra contribution to gravity, and we reject the possibility that it is a graviton, because such a contribution cannot remain undetected if it is added to gravity as we know it.

It would be wrong to think that this section is an advocacy of the "ninth gluon" or the double-copy correspondence, but if the end result of antimatter gravity experiments is different gravitational properties of atoms and antiatoms, we have to rethink everything.

Let us note that we still do not know if by its nature gravity (as other interactions) should be described by a quantum theory, but if yes, the quantum vacuum must contain fluctuations related to gravity.

In brief, there are a lot of unknowns and hence a lot of room for virtual gravitational dipoles within the quantum vacuum of the current Standard Model. In addition there are friendly extensions of the Standard Model which do not challenge its essence (for instance, addition of sterile neutrinos and the corresponding quantum vacuum fluctuations) and possibility that quantum vacuum contains fluctuations related to gravity if gravity is a quantum theory. We do not advocate any of these possibilities; the point is that the quantum vacuum as a source of gravity through the gravitational polarisation of the quantum vacuum is a so precious scenario that, despite respectable theoretical mainstream arguments against gravitational dipoles, we must stay open minded even (or especially) for the unlikely possibilities.

## 5.3. Gravitational Dipoles within the Framework of General Relativity

There are exciting and to some extent surprising possibilities, that gravitational dipoles (including virtual gravitational dipoles in the quantum vacuum) might exist within the framework of General Relativity

### 5.3.1. Gravitational Dipole Moment Related to Spin in General Relativity with Torsion

According to the Weak Equivalence Principle (telling us that inertial mass and gravitational charge is the same thing) negative gravitational charge and gravitational dipoles analogous to electric dipoles cannot exist within the framework of General relativity.

However, the existence of microscopic gravitational dipoles which are not similar to electric dipoles (for instance, to some extent they can be similar to magnetic dipoles existing without the existence of magnetic charges) remains possible. In fact, we already have a theoretical example of such a gravitational dipole in a simple extension of General Relativity; an exact solution [58] to the gravitational field equations *with torsion* shows that there is a dipole gravitational field, even though there is no negative gravitational charge. Let us underscore that the corresponding gravitational dipole moment $\mathbf{p}_g$ is related to the intrinsic spin $\mathbf{S}$ (which is a fundamentally quantum phenomenon); from the original paper [48] it is easy to get the following relation for the magnitude of $\mathbf{p}_g$:

$$p_g(S) = \frac{12}{A}\frac{GS^2}{c^4} \tag{30}$$

In the above equation $A$ is a constant surface which is not determined in the original paper; hence there is an unknown constant of proportionality $1/A$ (which in original paper was denoted by λ).

The Eq.(30) suggests that all virtual particles and antiparticles in the quantum vacuum (hence all quarks and leptons with spin $S = \hbar/2$, and gauge bosons with spin $S = \hbar$) are virtual gravitational dipoles, while Higgs Boson which is scalar boson with spin $S = 0$ is not a gravitational dipole. Hence, quantum vacuum fluctuations might be gravitational dipoles not because of the gravitational charges of the opposite sign, but because of spin and torsion.

In QV cosmology the magnitude of the gravitational dipole moment of a quantum vacuum fluctuation (composed of two gravitational charges of the opposite sign) [13, 15, 16] is equal to:

$$p_g = \frac{1}{2\pi}\frac{\hbar}{c} \tag{31}$$

While dipoles given by Eq.(30) and Eq.(31) have very different physical nature, assuming that they have the same magnitude, the gravitational field caused by one dipole would be indistinguishable from the gravitational field caused by the other dipole. Of course, the unknown constant A in Eq.(30) can be always chosen so that magnitude of two dipoles is the same (or has a very close value). However, such a mathematical choice of the constant $A$ is without any physical significance .

However, there is a surprising coincidence, which hopefully has physical significance. If we assume that constant $A$ is the surface of a "Planck sphere", i.e. $A = 4\pi l_p^2$ (where $l_p = \sqrt{\hbar G/c^3}$ is the Planck length), Eq.(30) reduces respectively to:

$$p_g\left(S = \frac{\hbar}{2}\right) = \frac{3}{4\pi}\frac{\hbar}{c}; \quad p_g(S = \hbar) = \frac{3}{\pi}\frac{\hbar}{c} \qquad (32)$$

It is astonishing and intriguing that dipoles related to intrinsic spin and dipoles related to the hypothetical existence of positive and negative gravitational charge in the quantum vacuum fluctuations have nearly the same magnitude.

In order to get the gravitational polarization of the quantum vacuum it is sufficient that only one of these two possibilities is realized in nature; in particular, in spite of different physical nature of gravitational dipoles, results obtained in [13-18] remain valid.

For better understanding, it is useful to underscore the following two things.

First, there is no fundamental reason, apart from simplicity, to assume (as it is assumed in General Relativity) that space-time is torsionless. The main feature of General Relativity is that mass-energy generates the curvature (and interacts with the curvature) of space-time, but the torsion of space-time is zero. Torsion is the most natural extension of General Relativity and apparently the best way to include effects related to intrinsic spin of particles (i.e. the best way of incorporating spin in a geometric description); it is quite possible that intrinsic spin generates the torsion (and interacts with the torsion) of space-time in an analogous way as mass-energy generates the curvature. Without guidance from experimental measurements, there are a number of open theoretical possibilities; we need the experimental tests at the intersection of quantum physics and General Relativity and hopefully we will have such tests in a visible future as for instance a recently proposed measurement [49] of general relativistic precession of intrinsic spin using a ferromagnetic gyroscope with unprecedented sensitivity.

Second. The general opinion is that in General relativity the impact of curvature is many orders of magnitude larger than the impact of torsion. For sure this is true if we assume a gravitationally featureless classical vacuum, but it is questionable (and perhaps wrong) in the quantum vacuum in which, because of spin, each virtual particle might be a virtual gravitational dipole with a significant impact of the gravitationally polarized quantum vacuum. Hence, if the polarized quantum vacuum is taken into account, the impact of torsion might be even larger than the impact of curvature.

*5.3.2. Gravitomagnetic Dipoles in Standard General Relativity*

Gravitomagnetism (this name suggests the existing analogy with electromagnetism) is an approximation of General Relativity at low velocities and weak gravitational fields. The key point is that gravitomagnetic equations (which follow from General Relativity) are analogous to electromagnetic Maxwell equations [38].

The main difference between Newtonian theory of gravity and Gravitomagnetism (which can be considered as a general relativistic extension of gravity) is that, in addition to the ordinary Newtonian gravitational field (described by Newtonian acceleration $\boldsymbol{g}_N$), there is a second, gravitomagnetic field $\boldsymbol{B}_{gm}$ caused by moving masses (hence, $\boldsymbol{B}_{gm}$ is the gravitational analogue of magnetic field $\boldsymbol{B}$ produced by the motion of electric-charges).

If we trust General Relativity we should trust the gravitomagnetism as well. However it is no longer a question of trusting. The Gravity B probe experiment [60, 61] has confirmed the existence of both, the gravitomagnetic field caused by the Earth's rotation, and the gravitomagnetic field caused by the Earth's orbiting around Sun.

From the point of view of the paradigm that quantum vacuum is a source of gravity (through the mechanism of the gravitational polarization) the most important result is that all rotating bodies (planets, stars…) produce a gravimagnetic field related to the angular momentum of the rotating body. Hence a rotating body is a gravitomagnetic dipole (analogous to magnetic dipole). Knowing this, it seems plausible that all particles with intrinsic spin are gravitomagnetic dipoles. Consequently, the quantum vacuum might be full of virtual gravitomagnetic dipoles which can be aligned (i.e. quantum vacuum can be polarised) by the gravitomagnetic field of bodies immersed in the quantum vacuum and moving through it.

## 6. Outlook

After surprising preliminary results from the ALPHA-g experiment, there is unexpected hope that antimatter gravity experiments could reveal new physics and profoundly change our understanding of gravity. In parallel, astronomical observations with a new generation of telescopes and instruments are challenging our best cosmological models, i.e. our current concepts of the content and evolution of the Universe.

We don't know if these two lines of research are related, and if they are somehow connected, what the connection is. Whatever is right or wrong, QV cosmology shows that, if our imagination is not switched off, we can find a common theoretical framework for the results of both lines of research.

It is absolutely necessary to speed up the gravitational experiments with positronium and muonium, because these experiments can provide a definitive answer (in the relatively near future) to the fundamental question of whether leptons and antileptons have equal or different gravitational charges. Indirectly, positronium and muonium experiments will help to understand experiments with antiatoms, which are more difficult to interpret (due to the complex structure of nucleons and antinucleons).

Astronomical observations appear to be on the verge of confirming a number of important challenges to our current model of the Universe. For example, we will soon have much better data on the surprisingly early formation of the first stars and galaxies, and we will know for sure whether what we call dark energy varies with time, and whether the Hubble tension is as real as it seems.

In the event that new physics is revealed by these two separate lines of research, we need to think about how to bring them together within a common theoretical framework. QV cosmology is the first attempt to do this, assuming in advance that new physics will emerge from both lines of research.

According to QV cosmology, the Universe is *nothing more* than a symbiosis of Standard Model matter (i.e. matter composed of quarks and leptons interacting through the exchange of gauge bosons) and the quantum vacuum.

The corresponding cosmological model is strictly based only on the Standard Model of particles and fields and General relativity, without any hypothetical content of the Universe (such as dark matter and dark energy) and without any modification of gravity. The exclusion of the hypothetical content and modification of gravity is made possible by a single working hypothesis that *quantum vacuum fluctuations are by their nature virtual gravitational dipoles*. In the case of random orientation of the gravitational dipoles, the quantum vacuum is not a source of gravity; however, the quantum vacuum is a source of gravity (with an effective gravitational charge density) due to the gravitational polarisation of the quantum vacuum by the Standard Model matter immersed in it.

QV cosmology is a general relativistic cosmology based on the cosmological principle and the standard cosmological field equations (as described in Section 3.1). There are only three cosmological fluids in QV cosmology. The first cosmological fluid is non-relativistic Standard Model matter, which is a pressureless fluid. The second fluid is the relativistic Standard Model matter, a positive pressure fluid. The third and most complex cosmological fluid is the quantum vacuum, which, depending on the epoch in the evolution of the Universe, can act as a fluid with positive, zero and negative pressure.

In QV cosmology, phenomena usually attributed to dark matter and dark energy and/or the modification of gravity are explained as local and global effects of the gravitationally polarised quantum vacuum. Galaxies do not have halos of dark matter, but halos of the polarised quantum vacuum, which locally explain the dynamics of the galaxy. Globally, all halos are a cosmological fluid in an epoch of negative pressure, causing the current accelerated expansion of the Universe. For the first time, the quantum vacuum is proposed as a substitute for both dark matter and dark energy.

As the Universe expands the negative pressure of the quantum vacuum decreases to zero. Due to the very large effective gravitational charge of the quantum vacuum, the expansion of the Universe would end and the contraction would begin. However the collapse would end at a relatively small but macroscopic size (hence, there is no singularity and no need for cosmic inflation). According to the QV cosmology, in a tiny fraction of the last second of the collapse, all the matter of our Universe would be converted into the antimatter of the next cycle of the Universe; the mechanism of this conversion is the enormous and superfast creation of particle-antiparticle pairs from the quantum vacuum by an extremely strong gravitational field in the final stage of the collapse. Thus, QV cosmology proposes an elegant explanation for the matter-antimatter asymmetry in the Universe: our cycle of the Universe is dominated by matter because the previous cycle was dominated by antimatter (and the next cycle would be dominated by antimatter again).

It is quite possible that the QV cosmology is wrong, but it is a cosmological model of rare beauty and elegance.

What if, despite the hopes raised by the ALPHA-g experiment, antimatter gravity experiments don't find signatures of new physics? Will that be the end of QV cosmology? Well, not really. In fact, antimatter gravity experiments cannot prove or disprove QV cosmology; they can only increase or decrease the plausibility of the theory. Put simply, if, for example, lepton-antilepton pairs are gravitational dipoles, that's no proof that other kinds of dipoles exist; if they're not dipoles, that's no proof that other kinds of dipoles don't exist. It would be great if new physics were revealed almost simultaneously in laboratory experiments and astronomical observations, but while highly desirable, it is not absolutely necessary. The most pragmatic approach is to speed up the development of QV cosmology and compare it with astronomical observations, which is the only way to confirm or reject the predictions of the theory.

## Appendix A: Matter-antimatter symmetric cosmology with different gravitational properties of atoms and antiatoms

As noted in the Introduction, there have been intriguing and imaginative attempts (Lattice Universe [8, 9] and Dirac-Milne cosmology [10, 11, 12]) to develop alternative cosmologies that challenge both the existence of hypothetical dark matter and dark energy and our belief that our universe is dominated by matter.

As a "prelude" to the Lattice Universe, it was shown that CPT symmetry and General relativity are compatible only if matter and antimatter[1] repel each other [28].

After this groundbreaking theoretical result, the next natural step was to see what would happen if there were equal amounts of matter and antimatter in the universe. As a simplified model, the universe was considered as a crystal lattice (analogous to an electrostatic lattice structure) with sites of the lattice alternately occupied by matter and antimatter domains. The key result is that the alternation of the unlike (positive and negative) gravitational charges in the universe produces a net accelerated expansion, despite the equal amounts of the two components. *Thus, there is an*

---

[1] A word of caution. The result obtained in reference [28] shouldn't be valid for composite systems, but only for fundamental particles and antiparticles. Therefore, if we trust the Standard Model of particles and fields, gravitational repulsion should exist between a quark and an antiquark, a lepton and an antilepton, but not between a proton and an antiproton, which are composite systems.

*accelerated expansion of the universe without dark energy*, which is a fascinating theoretical result whether nature works that way or not.

The gravitational interaction (actually repulsion) between atoms and antiatoms assumed in the lattice universe model seems incompatible with the ALPHA-g result. While there are arguments [29] that a reconciliation between experiment and lattice universe theory is possible, the tensions remain strong.

In Dirac-Milne cosmology, gravitational interactions are radically different from those in the lattice universe model. Atoms gravitationally attract both atoms and antiatoms, while antiatoms gravitationally repel both atoms and antiatoms; in other words, atoms are the source of universal gravitational attraction, while antiatoms are the source of universal gravitational repulsion.

A rigorous formulation of the theory with such complex gravitational interactions (and CPT symmetry violation) is only possible within the framework of a bimetric theory of gravity [12]. In simple terms, the postulated negative active gravitational charge of antimatter would repel other antimatter only if the passive gravitational charge of antimatter is positive. On the other hand, the positive active gravitational charge of matter repels antimatter only if the passive gravitational charge of antimatter is negative. Thus, the passive gravitational charge of antimatter must be positive in one case and negative in the other; in other words, the gravitational properties of matter and antimatter cannot be described by simply assigning a combination of signs to the three types of Newtonian masses (i.e., inertial mass, active gravitational mass, and passive gravitational mass). Instead, Dirac-Milne cosmology is formulated as a bimetric theory.

According to preliminary theoretical studies, Dirac-Milne cosmology is in astonishing agreement with observations; in particular, the theory passes relatively successfully the classical cosmological tests, such as primordial nucleosynthesis, Type Ia supernovae, and the cosmic microwave background.

Thus, both the Lattice universe and the bimetric Dirac-Milne cosmology have shown initial success in describing some of the observed features of the universe.

Note that the Lattice universe with huge and equal amounts of matter and antimatter is not empty, but it is gravitationally empty because the gravitational charges of the matter and antimatter domains cancel each other out; the total gravitational charge and the average gravitational charge density are zero. Therefore, according to cosmological equations (2) and (3), the lattice universe can be approximated by a single metric version of the Dirac-Milne universe (which is actually the metric of a gravitationally empty universe).

In short, the lattice universe in which CPT symmetry is valid can be approximated by a single metric Dirac-Milne cosmology, while if CPT is not respected, there is a bimetric Dirac-Milne cosmology. Therefore, although it remains to be verified, it seems plausible that some successes of the bimetric Dirac-Milne cosmology can also be interpreted as successes of the lattice universe. It is quite possible that single metric and bimetric Dirac-Milne universes give very similar results; this should be considered as an open question.

In any case, these alternative matter-antimatter symmetric cosmologies do not consider the quantum vacuum as a source of gravity; hence, they are radically different from the QV cosmology, which is based on the gravitational properties of the quantum vacuum enriched with virtual gravitational dipoles.

# Appendix B: Analytical Results and Open Questions in QV Cosmology

Our current understanding of the Universe (in fact the ΛCDM cosmology) is both, a fascinating scientific achievement, and the source of the greatest crisis in the history of physics. While ΛCDM cosmology is a very successful theory, we don't know what the nature of what we call dark matter is, what the nature of what we call dark energy is, what the nature of what we call cosmic inflation is (if it exists at all), what the cause of the challenging astronomical observations outlined in Section 4 is...

The absolutely astonishing success of ΛCDM cosmology has made the development of alternative cosmologies, including Quantum Vacuum cosmology, more difficult. The dominance of ΛCDM cosmology is well illustrated by the fact that thousands of scientists are working on its development, while only a few people are involved in QV cosmology. Nevertheless, QV cosmology has significant analytical results that are not widely known, and a number of plausible qualitative predictions that can be developed quantitatively.

## Appendix B.1: Analytical Solutions

*The first analytical solution* (see details of the solution in reference [8]) determines the gravitational field of a *point-like body immersed in the quantum vacuum*

We have absolutely no knowledge of the gravitational properties of the quantum vacuum, and to overcome our ignorance it is necessary to adopt an appropriate working hypothesis that fixes its gravitational properties. *In QV cosmology the working hypothesis is that quantum vacuum fluctuations* (for non-gravitational interactions described by the Standard Model of particles and fields) *behave as virtual gravitational dipoles*.

The solution of the corresponding fundamental equation (11) tells us that the total acceleration is a sum of the Newtonian (Einsteinian) acceleration and a tiny acceleration (always smaller than $g_{qvmax}$ which is of the order of $10^{-11}\, m/s^2$) due to the gravitational polarisation of the quantum vacuum caused by the gravitational field of the immersed body.

In Newton's theory of gravity and Einstein's theory of general relativity, the point-like body in question exists in a gravitationally featureless classical (non-quantum) vacuum; this is not surprising, since both theories were developed before the existence of the quantum vacuum was established. Obviously, if the quantum vacuum is a source of gravity, both theories are *incomplete*.

*The second analytical solution* within QV cosmology (see details of the solution in reference [36]) determines the gravitational field of two point-like bodies immersed in the quantum vacuum. Note that this second solution reveals, among other things, the existence of regions of the quantum vacuum with a negative effective gravitational charge density [36]; indeed, there is a region of negative effective gravitational charge surrounding the point where the ordinary gravitational fields of the bodies cancel each other out. This is just one example of complexity of the gravitational polarisation of the quantum vacuum. The islands of quantum vacuum with negative effective gravitational charge should also exist within the Solar System, but detecting them would be an extremely difficult task. In principle, apart from regions with a negative effective gravitational charge, the analytical solution for two point-like bodies immersed in the quantum vacuum can be tested in wide binaries and eventually in trans-Neptunian binaries [8, 36].

*Analytical solutions exist also for some particular distributions of matter* (see Chapter 3, section 3.3 of reference [10].

As an extreme but illuminating example of mass distribution, let's imagine an infinite plane of mass, with a constant surface mass density $\sigma$ (hence $\sigma$ is mass per unit area). Within the framework of the Newtonian gravity (using the Gauss's law of gravity), it is easy to show that the gravitational acceleration $\boldsymbol{g}_N$ caused by the plane, is directed towards the plane and has a constant magnitude $g_N = 2\pi G\sigma$ independent from distance $z$ from the plane. Hence, $g_N$ is a constant vector (with opposite direction on different sides of the plane).

If you see this result for the first time you may be surprised; the acceleration is independent of the distance from the plane! While this result is a consequence of the Newton's inverse square law, it is not "visible"; it is masked by a particular distribution of matter. This extreme example shows how important is the way in which a given mass is distributed.

Of course, the constant gravitational field $\boldsymbol{g}_N$ produces a constant gravitational polarization of the quantum vacuum, i.e., the vector $\boldsymbol{P}_g$ of the gravitational polarization density is also constant. Consequently, according to the fundamental equation $\rho_{qv} = -\boldsymbol{\nabla} \cdot \boldsymbol{P}_g$, the density $\rho_{qv}$ as a divergence of the constant vector $\boldsymbol{P}_g$ must be zero everywhere. Despite the existence of the gravitational polarization, the *quantum vacuum outside the plane is not a source of gravity*, because its effective gravitational charge is equal to zero. So if we imagine that a spherical distribution of Standard Model matter is somehow "deformed" into a planar-like distribution, an observer would measure the same amount of Standard Model matter, but different amounts of "dark matter", i.e. the effective gravitational charge of the quantum vacuum!

Between the examples of a point-like body and an infinite plane of matter, there is another illuminating example with an analytical solution, an infinitely long line of mass with constant mass per unit length, which produces a different pattern of gravitational polarisation of the quantum vacuum.

The conclusion is that the effects of the gravitational polarisation of the quantum vacuum depend significantly on the way in which Standard Model matter is distributed in the quantum vacuum.

Recall that astronomical observations suggest that the distribution of matter in the present-day universe has a web-like configuration called the Cosmic Web. Roughly speaking, the Cosmic Web has the following components: nodes (or knots), filaments, sheets (or walls) and voids. The gravitational polarisation caused by these very different components (galactic filaments are very elongated quasi one-dimensional structures, sheets are quasi planar structures, nodes are actually cluster regions) must be very different.

In conclusion, the concept of the gravitational polarisation of the quantum, caused by the well-known Standard Model matter immersed in it, is very simple and elegant but the result of the polarisation is very complex and despite the existence of very significant analytical solutions, QV cosmology must urgently turn to numerical methods and simulations.

## Appendix B.2: QV Cosmology versus MOND in a Galaxy

This part of the Appendix is a supplement to subsection 3.4.1. Interlude - Dark matter the first quantitative success of QV cosmology. The key point is that, in the case of a galaxy, QV cosmology produces results that are indistinguishable from those of MOND (which itself describes the dynamics of a galaxy as well as the dark matter hypothesis) at the current accuracy of astronomical measurements.

MOND starts with the *ad-hoc* assumption that, for a point-like source of gravity, the ratio of the total and Newtonian acceleration ($g_{tot}/g_N$) is a function of the ratio ($a_0/g_N$) of a universal acceleration $a_0$ and the Newtonian acceleration, i.e.

$$\frac{g_{tot}}{g_N} = f\left(\frac{a_0}{g_N}\right) > 1 \tag{B1}$$

Therefore the total acceleration is always greater than the Newtonian acceleration and can be easily calculated if the Newtonian acceleration and the interpolating function $f(a_0/g_N)$ are known.

Unfortunately the interpolating function $f(a_0/g_N)$ is not determined by the theory and MOND uses several ad-hoc interpolating functions, called the Simple, the Standard and RAR (the Radial Acceleration Relation) function. They are defined in Eq. (B2).

$$f_{smp}\left(\frac{a_0}{g_N}\right) = \frac{1}{2}\left(1 + \sqrt{4\frac{a_0}{g_N} + 1}\right), f_{std}\left(\frac{a_0}{g_N}\right) = \frac{1}{\sqrt{2}}\sqrt{1 + \sqrt{4\left(\frac{a_0}{g_N}\right)^2 + 1}}, \quad (B2)$$

$$f_{rar}\left(\frac{a_0}{g_N}\right) = \frac{1}{1 - e^{-\sqrt{g_N/a_0}}}$$

Of course, although these interpolating functions used in MOND look different their graphs must be nearly identical; otherwise they will predict different total accelerations.

We have now come to the first crucial test of QV cosmology. The ratio $g_{tot}/g_N$ i.e. the interpolating function $f(a_0/g_N)$, can be easily calculated using the Eq. (24) of QV cosmology. Thus, instead of ad hoc functions (B2), QV cosmology provides an analytic function (derived from the wild assumption that quantum vacuum fluctuations are virtual gravitational dipoles) that can be written (see reference [8]) as:

$$f_{qv}\left(\frac{a_0}{g_N}\right) \equiv \frac{g_{tot}}{g_N} = 1 + \alpha_1 \frac{a_0}{g_N} tanh\left(\alpha_2 \sqrt{\frac{g_N}{a_0}}\right); \; \alpha_1 \equiv \frac{g_{qvmax}}{a_0} \quad (B3)$$

Note that to ensure the appearance of $a_0$ in Eq. (B3), we have introduced dimensionless constants $\alpha_1 = g_{qvmax}/a_0 \approx 5/12$ and $\alpha_2 \approx 1/\alpha_1$. Figure A1 shows graphs of ad-hoc interpolating functions and our derived interpolating function (A3), which is the red line *between* two MOND's functions. More details and graphs can be found in section 6.1 of reference [8].

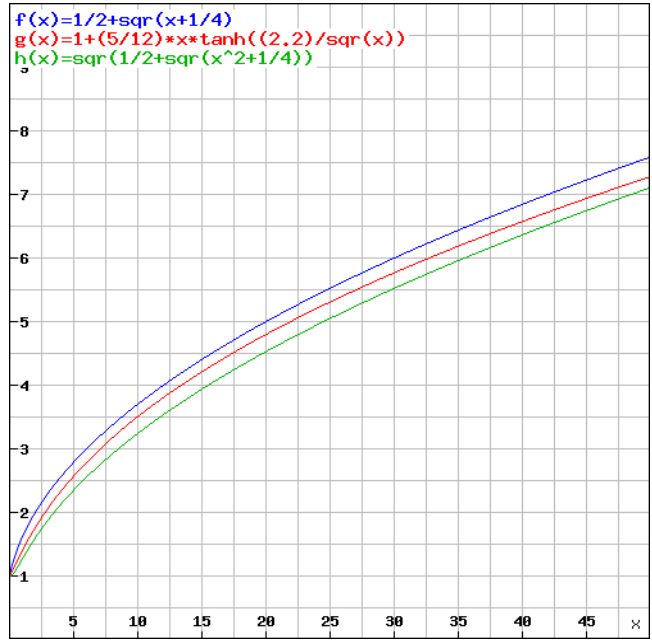

**Figure B1.** The simple and the standard interpolating functions in MOND (blue and green, respectively) compared with the prediction of the QV cosmology i.e. the polarized quantum vacuum (red line). In all functions: $x = a_0/g_N$.

In short, Eq. (B3) can reproduce all the remarkable results of MOND for galaxies. However, the underlying physics is radically different. In MOND, the total acceleration is greater than Newtonian because of a modification of the Newtonian law of gravity, whereas in QV cosmology the acceleration is greater than Newtonian because the quantum vacuum acts as a source of gravity; a hitherto neglected or, we might say, forgotten source of gravity. From the point of view of QV cosmology, MOND is a wrong physical theory, but correct in the mathematical description of two sources of gravity by a single non-Newtonian source of gravity.

Appendix B.3: QV Cosmology and "Dark Energy"

In the previous part of the appendix we saw that the gravitational polarisation of the quantum vacuum can explain galactic phenomena usually attributed to "dark matter" or the modification of gravity. The only way to get some insight into whether the quantum vacuum as a cosmological fluid can also explain phenomena usually attributed to "dark energy" is to consider a simplified model (for details see chapter 4 of reference [10]).

The main simplification in our model is that the Universe is composed of $N_h$ equal and *uniformly* distributed baryonic point-like gravitational sources and consequently of $N_h$ equal spherical halos of the polarised quantum vacuum. Of course, this is an oversimplification in today's Universe, which has a complex structure called the Cosmic Web. However, this is not far from the reality in the epoch when the content of the Universe was a rarefied gas of atoms (or ions) of hydrogen and helium; in that epoch, before the formation of stars and galaxies, the number of halos was approximately constant and equal to the number of atoms (ions). Note that instead of a model with $N_h$ equal atomic halos, it would be more realistic to consider a mixture of hydrogen and helium (hence two types of halo), but for simplicity we consider gas composed of only one type of atom.

According to this simplified model, the effective gravitational charge $M_{qvU}(R)$ of the Universe is:

$$M_{qvU}(R) = 4\pi P_{gmax} N_h^{1/3} R^2 \tanh\left(\frac{1}{N_h^{1/6}} \frac{1}{R} \sqrt{\frac{M_{SMU}}{4\pi P_{gmax}}}\right) \tanh\left(\frac{1}{N_h^{1/6}} \frac{\alpha_g}{R} \sqrt{\frac{M_{SMU}}{4\pi P_{gmax}}}\right) \quad \text{(B4)}$$

In the equation above $M_{SMU}$ is the Standard Model mass of the Universe. Equation (A4) tells us that the effective gravitational charge of the quantum vacuum in the Universe grows with the expansion of the Universe and tends asymptotically to $\alpha_g M_{SMU}$; hence each atom can have an effective gravitational charge about $\alpha_g$ times greater than its mass, as speculated in Section 3.4.3 and 4.2.

The corresponding equation of state is:

$$p_{qv}(R) = -c^2 \frac{\partial}{\partial V}[M_{qvU}(R)] = w_{qv}(R)\rho_{qv}(R)c^2 \quad \text{(B5)}$$

where, a variable equation of state parameter $w_{qv}(R)$ is given by

$$w_{qv}(R) = -\frac{2}{3}\left[1 - \frac{A_{sat}}{\sinh\left(2A_{sat}\frac{R_0}{R}\right)}\frac{R_0}{R} - \frac{\alpha_g A_{sat}}{\sinh\left(2n_{ran}A_{sat}\frac{R_0}{R}\right)}\frac{R_0}{R}\right] \quad \text{(B6)}$$

$$A_{sat} \equiv \frac{1}{N_h^{1/6}}\sqrt{\frac{\pi \rho_{SM0} R_0}{2 P_{gmax}}}$$

Note that Eq. (B6) is derived under the assumption that the number of halos $N_h$ is a constant (which is true for an expanding gas with a constant number of atoms, but not in general; for example, the number of halos decreases during the formation of the first stars); the $\rho_{SM0}$ denotes the present day density of Standard Model matter in the Universe. Note also that during the expansion, the quantum vacuum changes from a cosmological fluid with a significant negative pressure to a near-pressureless fluid (pressure tends asymptotically to zero).

Equation (B6), predicts that there is a period of accelerated expansion of the Universe filled with ionized gas of hydrogen and helium. According to cosmological equation (3), the accelerated expansion of the universe ends when the above equation of state parameter $w_{qv}(R)$ becomes greater than -1/3. The interesting question is what, in a universe expanding without acceleration, prevents the hydrogen-helium gas from collapsing. An amusing answer (speculated in Section 3.4.3) is that collapse is prevented by the onset of the formation of stars and galaxies.

Appendix B.4: QV Cosmology and the Cosmological Constant Problem

Consider a virtual electron-antielectron pair as a typical quantum-vacuum fluctuation. In fact, this fluctuation is an electric dipole with a total electric charge of zero. Since an electrically charged particle and its antiparticle have equal electric charges of opposite sign, we can immediately say that the total electric charge of the quantum vacuum is zero. The only reason why there is no electrical analogue of the cosmological constant problem is that quantum vacuum fluctuations are electric dipoles (or, as in the case of photons and gluons, they contain no electric charges). Now imagine that electrons and antielectrons have gravitational charges of the opposite sign; the corresponding quantum vacuum fluctuations are virtual gravitational dipoles with zero gravitational charge, and do not contribute to the gravitational charge of the quantum vacuum.

In summary, *the simplest possible solution to the cosmological constant problem is that **all** quantum vacuum fluctuations are virtual gravitational dipoles*. A positive and a negative gravitational charge within a fluctuation cancel each other out; consequently, the total gravitational charge of the quantum vacuum is zero, and (after many sophisticated and unsuccessful efforts) this may be a trivial solution to the cosmological constant problem.

Only quantum vacuum fluctuations that are not gravitational dipoles can be the source of the cosmological constant problem. For example, if antimatter gravity experiments show that quark-antiquark and lepton-antilepton pairs are gravitational dipoles, the solution to the cosmological constant problem would be reduced to understanding the gravitational properties of gauge bosons.

Appendix B.5: Open problems that are also crucial tests of QV cosmology

There are, of course, many open questions, but perhaps two of the most important are related to the comparison of the gravitational roles of dark matter and the gravitationally polarised quantum vacuum in the Cosmic Microwave Background power spectrum, and the formation of the first stars and galaxies.

Dark matter plays several crucial roles in ΛCDM cosmology. First, it is successfully used to complement baryonic matter in explaining the dynamics of galaxies and galaxy clusters. Second, the amount of Standard Model matter in the Universe is apparently insufficient for the initial efficient formation of stars and galaxies; it seems that without dark matter added to baryonic matter, the formation of the observed structure in the Universe is impossible. Third, we don't know how to fit the observed CMB power spectrum without dark matter.

An urgent task is therefore to see if stars and galaxies can efficiently form without dark matter, thanks to the effective gravitational charge of the quantum vacuum, which can be up to $\alpha_g$ times greater than the mass of baryonic matter (i.e. due to the effective gravitational charge of individual atoms, which can be up to $\alpha_g$ times greater than their mass). Another complementary task is to try to explain the CMB power spectrum without dark matter and to estimate the value of $\alpha_g$.

*Let us conclude this appendix by emphasising once again that the physical nature of gravitational dipoles in the quantum vacuum is a completely open question and, as seen in Section 5.3, is not necessarily related to antimatter gravity. Therefore, QV cosmology can survive any outcome of antimatter gravity experiments; in this context we mention the interesting work of Penner [52, 53]*

*who, inspired by QV cosmology, proposed a modification with gravitational dipoles attributed to virtual gravitons in the quantum vacuum.*